\documentclass[tradiabstract,twocolumn]{aa}
\usepackage{graphicx}
\usepackage{amsbsy}
\usepackage{natbib}
\usepackage{color}
\usepackage{txfonts}
\usepackage{url}
\newcommand{\vc}[1]{{\boldsymbol #1}}
\newcommand{\de}{\mathrm{d}}
\newcommand{\dpa}{\partial}

\newcommand{\nab}{\vc{\nabla}}

\DeclareMathSymbol{\varOmega}{\mathord}{letters}{"0A}
\DeclareMathSymbol{\varSigma}{\mathord}{letters}{"06}
\DeclareMathSymbol{\varPsi}{\mathord}{letters}{"09}
\DeclareMathSymbol{\varPhi}{\mathord}{letters}{"08}
\newcommand{\Eq}[1]{Eq.~(\ref{#1})}

\newcommand{\Sec}[1]{Sect.~\ref{#1}}

\newcommand{\Fig}[1]{Fig.~\ref{#1}}

\begin{document}

\title{High-resolution simulations of planetesimal formation\\in turbulent
protoplanetary discs}
\titlerunning{High-resolution simulations of planetesimal formation}

\author{Anders Johansen\inst{1,2}, Hubert Klahr\inst{3} \and Thomas
Henning\inst{3}}
\authorrunning{Johansen et al.} 

\offprints{\\A.\ Johansen (\email{anders@astro.lu.se})}

\institute{Lund Observatory, Box 43, 221 00 Lund, Sweden \and work partially
done at Leiden Observatory, Leiden University, P.O.\ Box 9513, 2300 RA Leiden,
The Netherlands \and Max-Planck-Institut f\"ur Astronomie, K\"onigstuhl 17,
69117 Heidelberg, Germany}

\abstract{We present high-resolution computer simulations of dust dynamics and
planetesimal formation in turbulence generated by the magnetorotational
instability. We show that the turbulent viscosity associated with
magnetorotational turbulence in a non-stratified shearing box increases when
going from $256^3$ to $512^3$ grid points in the presence of a weak imposed
magnetic field, yielding a turbulent viscosity of $\alpha\approx0.003$ at high
resolution. Particles representing approximately meter-sized boulders
concentrate in large-scale high-pressure regions in the simulation box.
The appearance of zonal flows and particle concentration in pressure
bumps is relatively similar at moderate ($256^3$) and high ($512^3$)
resolution. In the moderate-resolution simulation we activate particle
self-gravity at a time when there is little particle concentration, in contrast
with previous simulations where particle self-gravity was activated during a
concentration event. We observe that bound clumps form over the next ten
orbits, with initial birth masses of a few times the dwarf planet Ceres. At
high resolution we activate self-gravity during a particle concentration event,
leading to a burst of planetesimal formation, with clump masses ranging from a
significant fraction of to several times the mass of Ceres. We present a new
domain decomposition algorithm for particle-mesh schemes. Particles are spread
evenly among the processors and the local gas velocity field and assigned drag
forces are exchanged between a domain-decomposed mesh and discrete blocks of
particles. We obtain good load balancing on up to 4096 cores even in
simulations where particles sediment to the mid-plane and concentrate in
pressure bumps.
}
\maketitle
\section{Introduction}

The formation of km-scale planetesimals from dust particles involves a complex
interplay of physical processes, including most importantly collisional
sticking \citep{Weidenschilling1984,Weidenschilling1997,DullemondDominik2005},
the self-gravity of the particle mid-plane layer
\citep{Safronov1969,GoldreichWard1973,Sekiya1998,YoudinShu2002,SchraeplerHenning2004,Johansen+etal2007},
and the motion and structure of the turbulent protoplanetary disc gas
\citep{WeidenschillingCuzzi1993,Johansen+etal2006,Cuzzi+etal2008}.

In the initial growth stages micrometer-sized silicate monomers readily stick
to form larger dust aggregates \citep{Poppe+etal2000,BlumWurm2008}. Further
growth towards macroscopic sizes is hampered by collisional fragmentation and
bouncing \citep{Zsom+etal2010}, limiting the maximum particle size to a few cm
or less \citep[depending on the assumed velocity threshold for collisional
fragmentation, see][]{Brauer+etal2008a,Birnstiel+etal2009}. High-speed
collisions between small impactors and a large target constitutes a path to net
growth \citep{Wurm+etal2005}, but the transport of small particles away from
the mid-plane by turbulent diffusion limits the resulting growth rate
dramatically \citep{Johansen+etal2008}. Material properties are also important.
\cite{Wada+etal2009} demonstrated efficient sticking between ice aggregates
consisting of $0.1$ $\mu$m monomers at speeds up to 50 m/s.

Turbulence can play a positive role for growth by concentrating mm-sized
particles in convection cells \citep{KlahrHenning1997} and between small-scale
eddies \citep{Cuzzi+etal2008} occurring near the dissipative scale of the
turbulence. Larger m-sized particles pile up on large scales (i.e.\ larger than
the gas scale height) in long-lived geostrophic pressure bumps surrounded by
axisymmetric zonal flows \citep{Johansen+etal2009a}. In the model presented in
\cite{Johansen+etal2007} [hereafter referred to as J07], approximately
meter-sized particles settle to form a thin mid-plane layer in balance between
sedimentation and stirring by the gas which has developed turbulence
through the magnetorotational instability \citep{BalbusHawley1991}. Particles
then concentrate in nearly axisymmetric gas high-pressure regions which
appear spontaneously in the turbulent flow
\citep{FromangNelson2005,Johansen+etal2006,Lyra+etal2008a}, reaching local
column densities up to ten times the average. The passive concentration is
augmented as particles locally accelerate the gas towards the Keplerian speed,
which leads to accumulation of particles drifting rapidly in from exterior
orbits \citep[a manifestation of the streaming instability
of][]{YoudinGoodman2005}. The gravitational attraction between the particles in
the overdense regions becomes high enough to initiate first a slow radial
contraction, and as the local mass density becomes comparable to the Roche
density, a full non-axisymmetric collapse to form gravitationally bound clumps
with masses comparable to the 950-km-diameter dwarf planet Ceres ($M_{\rm
Ceres}\approx9.4\times10^{20}\,{\rm kg}$). Such large planetesimal birth sizes
are in agreement with constraints from the current observed size distribution
of the asteroid belt \citep{Morbidelli+etal2009} and Neptune Trojans
\citep{SheppardTrujillo2010}.

Some of the open questions related to this picture of planetesimal
formation is to what degree the results of \cite{Johansen+etal2007} are
affected by the fact that self-gravity was turned on after particles had
concentrated in a pressure bumps and how the emergence and amplitude of
pressure bumps are affected by numerical resolution. In this paper we present
high-resolution and long-time-integration simulations of planetesimal formation
in turbulence caused by the magnetorotational instability (MRI). We find that
the large-scale geostrophic pressure bumps that are responsible for particle
concentration are sustained when going from moderate ($256^3$) to high
($512^3$) resolution.  Particle concentration in these pressure bumps is also
relatively independent on resolution. We present a long-time-integration
simulation performed at moderate resolution ($256^3$) where particles and
self-gravity are started at the same time, in contrast to earlier simulations
where self-gravity was not turned on until a strong concentration event
occurred (J07). We also study the initial burst of planetesimal formation at
$512^3$ resolution. We present evidence for collisions between
gravitationally bound clumps, observed at both moderate and high
resolution, and indications that the Initial Mass Function of gravitationally
bound clumps involves masses ranging from a significant fraction of to several
times the mass mass of Ceres.
We point out that the physical nature of the collisions is unclear, since
our numerical algorithm does not allow clumps to contract below the grid size.
Gravitational scattering and binary formation are other possible outcomes of
the close encounters, in case of resolved dynamics. Finding the Initial Mass
Function of planetesimals forming from the gravitationally bound clumps will
ultimately require an improved algorithm for the dynamics and interaction
of bound clumps as well as the inclusion of particle shattering and
coagulation during the gravitational contraction.

The paper is organised as follows. In \Sec{s:equations} we describe the
dynamical equations for gas and particles. \Sec{s:code} contains descriptions
of a number of improvements made to the Pencil Code in order to be able to
perform particle-mesh simulations at up to at least 4096 cores. In
\Sec{s:parameters} we explain the choice of simulation parameters. The
evolution of gas turbulence and large-scale pressure bumps is analysed in
\Sec{s:gas}. Particle concentration in simulations with no self-gravity is
described in \Sec{s:particles}. Simulations including particle self-gravity are
presented in \Sec{s:sgmod} ($256^3$ resolution) and \Sec{s:sghigh} ($512^3$
resolution). We summarise the paper and discuss the implications of our results
in \Sec{s:discussion}.

\section{Dynamical equations}
\label{s:equations}

We perform simulations solving the standard shearing box MHD/drag
force/self-gravity equations for gas defined on a fixed grid and solid
particles evolved as numerical superparticles. We use the Pencil Code, a sixth
order spatial and third order temporal symmetric finite difference
code\footnote{See \url{http://code.google.com/p/pencil-code/}.}.

We model the dynamics of a protoplanetary disc in the shearing box
approximation. The coordinate frame rotates at the Keplerian frequency
$\varOmega$ at an arbitrary distance $r_0$ from the central star. The axes are
oriented such that the $x$ points radially away from the central gravity
source, $y$ points along the Keplerian flow, while $z$ points vertically out of
the plane.

\subsection{Gas velocity}

The equation of motion for the gas velocity $\vc{u}$ relative to the Keplerian
flow is
\begin{eqnarray}\label{eq:eqmot}
  \frac{\dpa \vc{u}}{\dpa t} +
    (\vc{u} \cdot \nab) \vc{u} + u_y^{(0)} \frac{\dpa \vc{u}}{\dpa y} &=& 2
    \varOmega u_y \vc{e}_x - \frac{1}{2} \varOmega u_x \vc{e}_y + 2 \varOmega
    \Delta v \vc{e}_x \nonumber \\
    & & \hspace{-2.5cm} + \frac{1}{\rho} \vc{J} \times
    (\vc{B}+B_0\hat{\vc{z}}) - \frac{1}{\rho}
    \nab P -\frac{\rho_{\rm p}/\rho_{\rm g}}{\tau_{\rm f}}
    (\vc{u}-\overline{\vc{v}}) + \vc{f}_\nu(\vc{u}) \, .
\end{eqnarray}
The left hand side includes advection both by the velocity field $\vc{u}$
itself and by the linearised Keplerian flow $u_y^{(0)}=-(3/2)\varOmega x$. The
first two terms on the right hand side represent the Coriolis force in the $x$-
and $y$-directions, modified in the $y$-component by the radial advection of
the Keplerian flow, $\dot{u}_y=-u_x \dpa u_y^{(0)}/\dpa x$. The third term
mimics a global radial pressure gradient which reduces the orbital speed of the
gas by the positive amount $\Delta v$. The fourth and fifth terms in
\Eq{eq:eqmot} are the Lorentz and pressure gradient forces. The current density
is calculated from Amp\`ere's law $\mu_0\vc{J}=\nab\times\vc{B}$. The Lorentz
force is modified to take into account a mean vertical field component of
strength $B_0$. The sixth term is a drag force term is described in Sect.\
\ref{s:drag}.

The high-order numerical scheme of the Pencil Code has very little numerical
dissipation from time-stepping the advection term \citep{Brandenburg2003}, so
we add explicit viscosity through the term $\vc{f}_\nu(\vc{u})$ in
\Eq{eq:eqmot}. We use sixth order hyperviscosity with a constant dynamic
viscosity $\mu_3=\nu_3\rho$,
\begin{equation}
  \vc{f}_\nu = \frac{\mu_3}{\rho_{\rm g}} \nabla^6 \vc{u} \, .
\end{equation}
This form of the viscosity conserves momentum. The $\nabla^6$ operator is
defined as $\nabla^6=\dpa^6/\dpa x^6 + \dpa^6/\dpa y^6 + \dpa^6/\dpa z^6$. It
was shown by \cite{Johansen+etal2009a} that hyperviscosity simulations show
zonal flows and pressure bumps very similar to simulations using Navier-Stokes
viscosity.

\subsection{Gas density}

The continuity equation for the gas density $\rho$ is
\begin{equation}
  \frac{\dpa \rho}{\dpa t} + (\vc{u} \cdot \nab) \rho + u_y^{(0)} \frac{\dpa
  \rho }{\dpa y} = -\rho \nab \cdot \vc{u} + f_D(\rho) \, .
\end{equation}
The diffusion term is defined as
\begin{equation}
  f_{\rm D} = D_3 \nabla^6 \rho \, ,
\end{equation}
where $D_3$ is the hyperdiffusion coefficient necessary to suppress Nyquist
scale wiggles arising in regions where the spatial density variation is high.
We adopt an isothermal equation of state with pressure $P=c_{\rm s}^2 \rho$ and
(constant) sound speed $c_{\rm s}$.

\subsection{Induction equation}

The induction equation for the magnetic vector potential $\vc{A}$ is
\citep[see][for details]{Brandenburg+etal1995}
\begin{equation}
  \frac{\dpa \vc{A}}{\dpa t} + u_y^{(0)} \frac{\dpa \vc{A}}{\dpa y} =
  \vc{u}\times(\vc{B}+B_0\hat{\vc{z}}) + \frac{3}{2} \varOmega A_y
  \hat{\vc{x}} + \vc{f}_\eta(\vc{A}) \, .
\end{equation}
The resistivity term is
\begin{equation}
  \vc{f}_\eta = \eta_3 \nabla^6 \vc{A} \, ,
\end{equation}
where $\eta_3$ is the hyperresistivity coefficient. The magnetic field is
calculated from $\vc{B}=\nab\times\vc{A}$.

\subsection{Particles and drag force scheme}
\label{s:drag}

The dust component is treated as a number of individual superparticles, the
position $\vc{x}$ and velocity $\vc{v}$ of each evolved according to
\begin{eqnarray}\label{eq:eqmotpar}
  \frac{\de \vc{x}}{\de t} &=& \vc{v} - \frac{3}{2} \varOmega x
  \vc{e}_y \, , \\
  \frac{\de \vc{v}}{\de t} &=& 
    2 \varOmega v_y \vc{e}_x - \frac{1}{2} \varOmega v_x \vc{e}_y + \nab
    \varPhi - \frac{1}{\tau_{\rm f}} (\vc{v}-\overline{\vc{u}}) \, .
    \label{eq:dvpdt}
\end{eqnarray}
The particles feel no pressure or Lorentz forces, but are subjected to the
gravitational potential $\varPhi$ of their combined mass. Particle collisions
are taken into account as well (see Sect.\ \ref{s:coll} below).

Two-way drag forces between gas defined on a fixed grid and Lagrangian
particles are calculated through a particle-mesh method \citep[see][for
details]{YoudinJohansen2007}. First the gas velocity field is interpolated to
the position of a particle, using second order spline interpolation. The drag
force on the particle is then trivial to calculate. To ensure momentum
conservation we then take the drag force and add it with the opposite sign
among the 27 nearest grid points, using the Triangular Shaped Cloud scheme to
ensure momentum conservation in the assignment \citep{HockneyEastwood1981}.

\subsection{Units}

All simulations are run with natural units, meaning that we set $c_{\rm
s}=\varOmega=\mu_0=\rho_0=1$. Here $\rho_0$ represents the mid-plane gas
density, which in our unstratified simulations is the same as the mean density
in the box. The time and velocity units are thus $[t]=\varOmega^{-1}$ and
$[v]=c_{\rm s}$. The derived unit of the length is the scale height $H\equiv
c_{\rm s}/\varOmega=[l]$. The magnetic field unit is $[B]=c_{\rm
s}\sqrt{\mu_0\rho_0}$.

\subsection{Self-gravity}

The gravitational attraction between the particles is calculated by first
assigning the particle mass density $\rho_{\rm p}$ on the grid, using the
Triangular Shaped Cloud scheme described above. Then the gravitational
potential $\varPhi$ at the grid points is found by inverting the Poisson
equation
\begin{equation}
  \nabla^2 \varPhi = 4 \pi G \rho_{\rm p}
\end{equation}
using a Fast Fourier Transform (FFT) method (see online supplement of J07).
Finally the self-potential of the particles is interpolated to the positions of
the particles and the acceleration added to the particle equation of motion
(Eq.\ \ref{eq:dvpdt}). We define the strength of self-gravity through the
dimensionless parameter
\begin{equation}
  \tilde{G} = \frac{4 \pi G}{\varOmega^2/\rho_0} \, ,
\end{equation}
where $\rho_0$ is the gas density in the mid-plane. This is practical since all
simulations are run with $\varOmega=\rho_0=c_{\rm s}=1$. Using $\varSigma_{\rm
g}=\sqrt{2\pi} \rho_0 H$ for the gas column density, we obtain a connection
between the dimensionless $\tilde{G}$ and the relevant dimensional parameters
of the box,
\begin{equation}
  \label{eq:Sigma}
  \varSigma_{\rm g} = 3600 \tilde{G}\,{\rm g\,cm^{-2}}
  \left( \frac{H/r}{0.05} \right) \left( \frac{M_\star}{\rm M_\odot}
  \right) \left( \frac{r}{5\,{\rm AU}} \right)^{-2} \, .
\end{equation}
We assume a standard ratio of particle to gas column densities of 0.01. The
self-gravity of the gas is ignored in both the Poisson equation and the gas
equation of motion.

\subsection{Collisions}
\label{s:coll}

Particle collisions become important inside dense particle clumps. In J07 the
effect of particle collisions was included in a rather crude way by reducing
the relative rms speed of particles inside a grid cell on the collisional
time-scale, to mimic collisional cooling. Recently \cite{Rein+etal2010} found
that the inclusion of particle collisions suppresses condensation of small
scale clumps from the turbulent flow and favours the formation of larger
structures.

\cite{Rein+etal2010} claimed that the lack of collisions is an inherent flaw in
the superparticle approach. However, \cite{LithwickChiang2007} presented
a scheme for the collisional evolution of particle rings whereby the collision
between two particles occupying the same grid cell is determined by drawing a
random number (to determine whether the two particles are at the same vertical
position). We have extended this algorithm to model collisions between
superparticles based on a Monte Carlo algorithm. We obtain the correct
collision frequency by letting nearby superparticle pairs collide on the
average once per collisional time-scale of the swarms of physical particles
represented by each superparticle.

We have implemented the superparticle collision algorithm in the Pencil Code and
will present detailed numerical tests that show its validity in a paper in
preparation (Johansen, Youdin, \& Lithwick, in preparation). The algorithm
gives each particle a chance to interact with all other particles in the same
grid cell.  The characteristic time-scale $\tau_{\rm coll}^{(ij)}=1/(n_j
\sigma_{ij} \delta v_{ij})$ for a representative particle in the swarm of
superparticle $i$ to collide with any particle from the swarm represented by
superparticle $j$ is calculated by considering the number density $n_j$
represented by each superparticle, the collisional cross section $\sigma_{ij}$
of two swarm particles, and the relative speed $\delta v_{ij}$ of the two
superparticles.  For each possible collision a random number $P$ is chosen. If
$P$ is smaller than $\delta t/\tau_{\rm coll}$, where $\delta t$ is the
time-step set by magnetohydrodynamics, then the two particles collide. The
collision outcome is determined as if the two superparticles were actual
particles with radii large enough to touch each other. By solving for momentum
conservation and energy conservation, with the possibility for inelastic
collisions to dissipate kinetic energy to heat and deformation, the two
colliding particles acquire their new velocity vectors instantaneously.

All simulations include collisions with a coefficient of restitution of
$\epsilon=0.3$ \citep[e.g][]{BlumMuench1993}, meaning that each collision leads
to the dissipation of approximately 90\% of the relative kinetic energy to
deformation and heating of the colliding boulders. We include particle
collisions here to obtain a more complete physical modelling. Detailed tests
and analysis of the effect of particle collisions on clumping and gravitational
collapse will appear in a future publication (Johansen, Youdin, \& Lithwick, in
preparation).

\section{Computing resources and code optimisation}
\label{s:code}

For this project we were kindly granted access to 4096 cores on the
``Jugene'' Blue Gene/P system at the J\"ulich Supercomputing Centre (JSC) for a
total of five months. Each core at the BlueGene/P has a clock speed of around
800 MHz.  The use of the Pencil Code on several thousand cores required both
trivial and more fundamental changes to the code. We describe these technical
improvements in detail in this appendix.

In the following {\tt nxgrid}, {\tt nygrid}, {\tt nzgrid} refer to the
full grid dimension of the problem. We denote the processor number by {\tt
ncpus} and the directional processor numbers as {\tt nprocx}, {\tt nprocy},
{\tt nprocz}.

\subsection{Changes made to the Pencil Code}

\subsubsection{Memory optimisation}
\label{s:memory}

\begin{figure}
  \begin{center}
    \includegraphics[width=8.7cm]{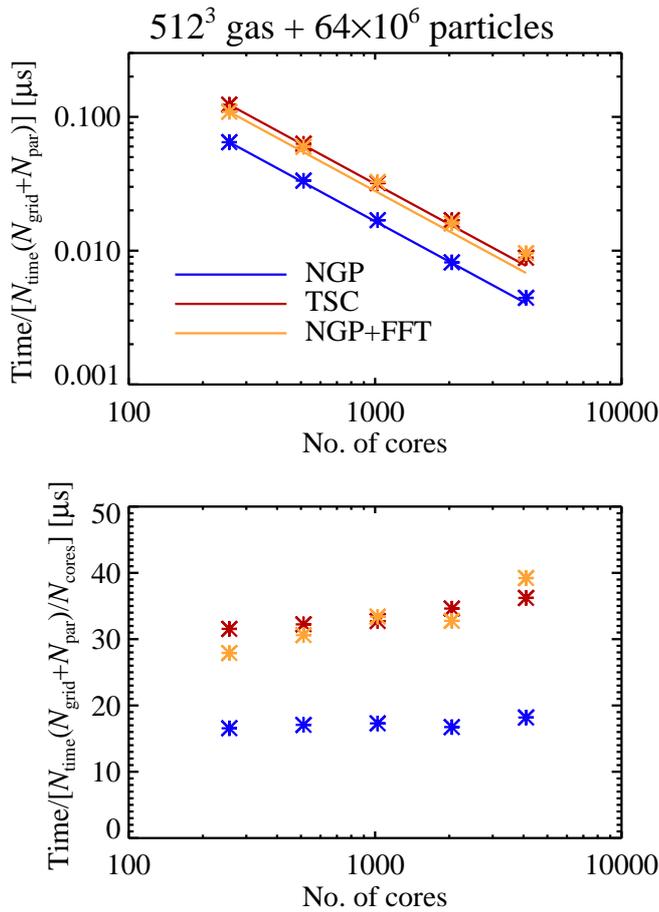}
  \end{center}
  \caption{Scaling test for particle-mesh problem with $512^3$ grid cells and
  $64\times10^6$ particles. The particles are distributed evenly over the grid,
  so that each core has the same number of particles. The inverse code speed is
  normalised by the number of time-steps and by either the total number of grid
  points and particles (top panel) or by the number of grid points and
  particles per core (bottom panel).}
  \label{f:timings}
\end{figure}
We had to remove several uses of {\it global} arrays, i.e. 2-D or 3-D
arrays of linear size equal to the full grid. This mostly affected certain
special initial conditions and boundary conditions. An additional problem was
the use of an array of size {\tt (ncpus,ncpus)} in the particle communication.
This array was replaced by a 1-D array with no problems.

The runtime calculation of 2-D averages (e.g.\ gas and particle column
density) was done in such a way that the whole {\tt (nxgrid,nygrid)} array was
collected at the root processor in an array of size {\tt
(nxgrid,nygrid,ncpus)}, before appending a chunk of size {\tt (nxgrid,nygrid)}
to an output file. The storage array, used for programming convenience in
collecting the 2-D average from all the relevant cores, became excessively
large at high resolution and processor numbers, and we abandoned the previous
method in favour of saving chunks of the column density array into separate
files, each maintained by the root processors in the $z$-direction. A similar
method was implemented for $y$-averages and $x$-averages.
\begin{figure}
  \includegraphics[width=8.7cm]{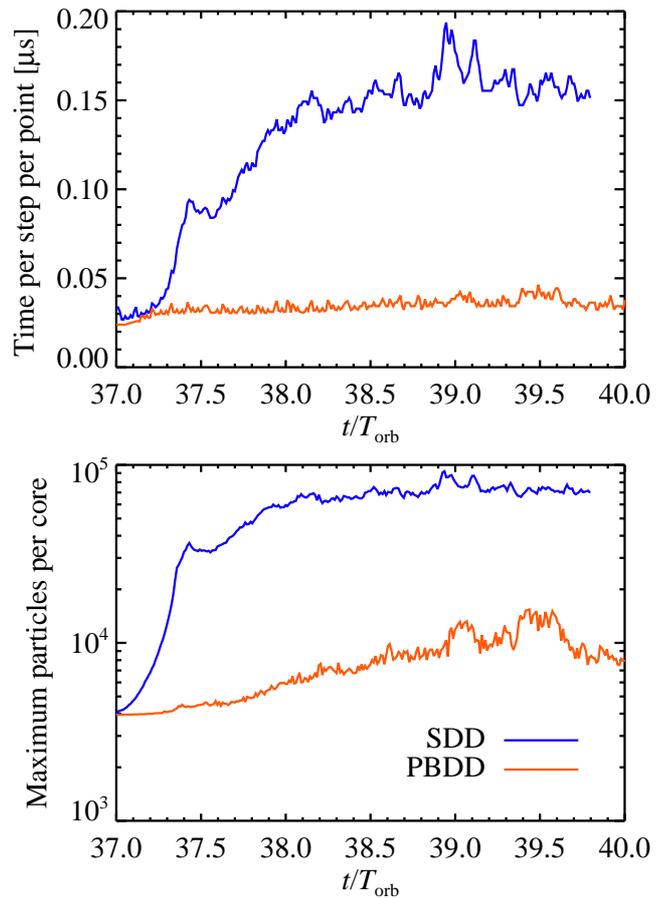}
  \caption{Code speed as a function of simulation time (top panel) and
  maximum particle number on any core (bottom panel) for $256^3$ resolution on
  2048 cores. Standard domain decomposition (SDD) quickly becomes unbalanced
  with particles and achieves only the speed of the particle-laden mid-plane
  cores.  With the Particle Block Domain Decomposition (PBDD) scheme the speed
  stays close to its optimal value, and the particle number per core (bottom
  panel) does not rise much beyond $10^4$.}
  \label{f:speed_t}
\end{figure}

The above-mentioned global arrays had been used in the code for
programming convenience and did not imply excessive memory usage at moderate
resolution and processor numbers. Purging those arrays in favour of loops or
smaller 1-D arrays was relatively straight-forward.

\subsubsection{Particle migration}
\label{s:parmig}

At the end of a sub-time-step each processor checks if any particles have
left its spatial domain. Information about the number of migrating particles,
and the processors that they will migrate into, is collected at each processor.
The Pencil Code would then let all processors exchange migrating particles with
all other processors. In practice particles would of course only migrate to
neighbouring processors. However, at processor numbers of 512 or higher, the
communication load associated with each processor telling all other processors
how many particles it wanted to send (mostly zero) was so high that it
dominated over both the MHD and the particle evolution equations. Since
particles in practice only migrate to the neighbouring processors any way, we
solved this problem by letting the processors only communicate the number of
migrating particles to the immediate neighbours. Shear-periodic boundary
conditions require a (simple) algorithm to determine the three neighbouring
processors over the shearing boundary in the beginning of each sub-time-step.

\subsection{Timings}

With the changes described in Sect.\ \ref{s:memory} and Sect.
\ref{s:parmig} the Pencil Code can be run with gas and particles efficiently at
several thousand processors, provided that the particles are well-mixed with
the gas.

\begin{table*}[!t]
  \caption{Simulation parameters in natural units.}
  \begin{center}
    \begin{tabular}{lccccccccccc}
      \hline
      \hline
      Run & $L_x \times L_y \times L_z$
          & $N_x \times N_y \times N_z$
          & $\nu_3=D_3=\eta_3$
          & $B_0$
          & $\beta$
          & $\Delta v$
          & $\tilde{G}$
          & $\Delta t$
          & $t_{\rm par}$
          & $t_{\rm grav,1}$
          & $t_{\rm grav,2}$ \\
      \hline
      M1 & $(1.32)^3$ & $256^3$ & $2\times10^{-14}$ & $0.0015$ & $9\times10^5$
         & $0.05$ & N/A   & $40.0$ & N/A    & N/A    & N/A \\
      M2 & $(1.32)^3$ & $256^3$ & $2\times10^{-14}$ & $0.003$  & $2\times10^5$
         & $0.05$ & $0.5$ & $40.0$ & $20.0$ & $20.0$ & $33.0$ \\
      H  & $(1.32)^3$ & $512^3$ & $7\times10^{-16}$ & $0.0015$  & $9\times10^5$
         & $0.05$ & $0.5$ & $40.0$ & $32.0$ & $35.0$ & $37.0$ \\
      \hline
    \end{tabular}
  \end{center}
  The first column gives the name of the simulation, while the box size
  and the grid resolution are given in the following two columns. The next
  column gives the hyperdiffusivity coefficients. The next two columns give the
  mean vertical magnetic field and the corresponding plasma-$\beta$. The next
  column gives the sub-Keplerian velocity difference. The following column
  shows the particle self-gravity parameter $\tilde{G}$. The last four columns
  give the total simulation time, the time when particles were released, and
  the times at which self-gravity was started and self-gravity simulations were
  stopped.
  \label{t:parameters}
\end{table*}
In \Fig{f:timings} we show timings for a test problem with $512^3$ grid
cells and $64\times10^6$ particles. The particles are distributed evenly over
the grid, avoiding load balancing challenges described below. We evolve gas and
particles for 1000 time-steps, the gas and particles subject to standard
shearing box hydrodynamics and two-way drag forces. The lines show various drag
force schemes -- NGP corresponding to Nearest Grid Point and TSC to Triangular
Shaped Cloud \citep{HockneyEastwood1981}. We achieve near perfect scaling up to
4096 cores. Including self-gravity by a Fast Fourier Transform (FFT) method the
code slows down by approximately a factor of two, but the scaling is only
marginally worse than optimal, with less than 50\% slowdown at 4096 cores. This
must be seen in the light of the fact that 3-D FFTs involve several
transpositions of the global density array, each transposition requiring the
majority of grid points to be communicated to another processor (see online
supplement of J07).

\subsection{Particle parallelisation}

At high resolution it becomes increasingly more important to parallelise
efficiently along at least two directions. In earlier publications we had run
$256^3$ simulations at 64 processors, with the domain decomposition {\tt
nprocy=32} and {\tt nprocz=2} (J07). Using two processors along the
$z$-direction exploits the intrinsic mid-plane symmetry of the particle
distribution, while the Keplerian shear suppresses most particle density
variation in the azimuthal direction, so that any processor has approximately
the same number of particles.

However, at higher resolution we need to either have {\tt nprocz>2} or
{\tt nprocx>1}, both of which is subject to particle clumping (from either
sedimentation or from radial concentrations). This would in some cases slow
down the code by a factor of 8-10. We therefore developed an improved particle
parallelisation, which we denote {\it Particle Block Domain Decomposition}
(PBDD). This new algorithm is described in detail in the following subsection.

\subsubsection{Particle Block Domain Decomposition}

The steps in Particle Block Domain Decomposition scheme are as follows:
\begin{enumerate}
  \item The fixed mesh points are domain-decomposed in the usual way (with {\tt
  ncpus}={\tt nprocx}$\times${\tt nprocy}$\times${\tt nprocz}).
  \item Particles on each processor are counted in {\it bricks} of size {\tt
  nbx}$\times${\tt nby}$\times${\tt nbz} (typically {\tt nbx}$=${\tt
  nby}$=${\tt nbz}$=${\tt 4}).
  \item Bricks are distributed among the processors so that each processor has
  approximately the same number of particles
  \item Adopted bricks are referred to as {\it blocks}.
  \item The Pencil Code uses a third order Runge-Kutta time-stepping scheme. In
  the beginning of each sub-time-step particles are counted in blocks and the
  block counts communicated to the bricks on the parent processors. The
  particle density assigned to ghost cells is folded across the grid, and the
  final particle density (defined on the bricks) is communicated back to the
  adopted blocks. This step is necessary because the drag force time-step
  depends on the particle density, and each particle assigns density not just
  to the nearest grid point, but also to the neighbouring grid points.
  \item In the beginning of each sub-time-step the gas density and gas velocity
  field is communicated from the main grid to the adopted particle blocks.
  \item Drag forces are added to particles and back to the gas grid points in
  the adopted blocks. This partition aims at load balancing the calculation of
  drag forces.
  \item At the end of each sub-time-step the drag force contribution to the gas
  velocity field is communicated from the adopted blocks back to the main grid.
\end{enumerate}

To illustrate the advantage of this scheme we show in \Fig{f:speed_t} the
instantaneous code speed for a problem where the particles have sedimented to
the mid-plane of the disc. The grid resolution is $256^3$ and we run on 2048
cores, with 64 cores in the $y$-direction $32$ cores in the $z$-direction. The
blue (black) line shows the results of running with standard domain
decomposition, while the orange (grey) line shows the speed with the improved
Particle Block Domain Decomposition scheme. Due to the concentration of
particles in the mid-plane the standard domain decomposition leaves many cores
with few or no particles, giving poor load balancing. This problem is
alleviated by the use of the PBDD scheme (orange/grey line).

PBDD works well as long as the single blocks do not achieve higher
particle density than the optimally distributed particle number ${\tt
npar}/{\tt ncpus}$. In the case of strong clumping, e.g.\ due to self-gravity,
the scheme is no longer as efficient. In such extreme cases one should ideally
limit the local particle number in clumps by using sink particles.

\section{Simulation parameters}
\label{s:parameters}

The main simulations of the paper focus on the dynamics and self-gravity
of solid particles moving in a gas flow which has developed turbulence through
the magnetorotational instability \citep{BalbusHawley1991}.
We have performed two moderate-resolution simulations (with $256^3$ grid points
and $8\times10^6$ particles) and one high-resolution simulation ($512^3$ grid
points and $64\times10^6$ particles). Simulation parameters are given in Table
\ref{t:parameters}. We use a cubic box of dimensions $(1.32 H)^3$.

Note that we use a sub-Keplerian speed difference of $\Delta v=0.05$ which is
higher than $\Delta v=0.02$ presented in the main paper of J07. The ability of
pressure bumps to trap particles is generally reduced with increasing $\Delta
v$ (see online supplementary information for J07). Particle clumping by
streaming instabilities also becomes less efficient as $\Delta v$ is increased
\citep[J07,][]{BaiStone2010c}. Estimates of the radial pressure support in
discs can be extracted from models of column density and temperature
structure. A gas parcel orbiting at a radial distance $r$ from the star, where
the disc aspect ratio is $H/r$ and the mid-plane radial pressure gradient is
$\de \ln P/\de \ln r$, orbits at a sub-Keplerian speed $v=v_{\rm K}-\Delta v$.
The speed reduction $\Delta v$ is given by
\begin{equation}
  \frac{\Delta v}{c_{\rm s}} = -\frac{1}{2} \frac{H}{r} \frac{\de \ln P}{\de
  \ln r} \, .
\end{equation}
In the Minimum Mass Solar Nebula of \cite{Hayashi1981} $\de \ln P/\de \ln
r=-3.25$ in the mid-plane \citep[e.g.][]{Youdin2008}. The resulting scaling of
the sub-Keplerian speed with the orbital distance is
\begin{equation}
  \frac{\Delta v}{c_{\rm s}} \approx 0.05 \left( \frac{r}{\rm AU} \right)^{1/4}
  \, .
\end{equation}
The slightly colder disc model used by \cite{Brauer+etal2008a} yields instead
\begin{equation}
  \frac{\Delta v}{c_{\rm s}} \approx 0.04 \left( \frac{r}{\rm AU} \right)^{1/4}
  \, .
\end{equation}
In more complex disc models the pressure profile is changed e.g.\ in interfaces
between regions of weak and strong turbulence \citep{Lyra+etal2008b}. We use
throughout this paper a fixed value of $\Delta v/c_{\rm s}=0.05$.
  
Another difference from the simulations of J07 is that the turbulent viscosity
of the gas flow is around 2-3 times higher, because of the increased turbulent
viscosity when going from $256^3$ to $512^3$ (see \Sec{s:gas}). Therefore we
had to use a stronger gravity in this paper, $\tilde{G}=0.5$ compared to
$\tilde{G}=0.2$ in J07, to see bound particle clumps (planetesimals) forming.
We discuss the implications of using a higher disc mass further in our
conclusions.

In all simulations we divide the particle component into four size bins, with
friction time $\varOmega\tau_{\rm f}=0.25,0.50,0.75,1.00$, respectively. The
particles drift radially because of the headwind from the gas orbiting at
velocity $u_y=-\Delta v$ relative to the Keplerian flow
\citep{Weidenschilling1977a}. It is important to consider a distribution of
particle sizes, since the dependence of the radial drift on the particle sizes
can have a negative impact on the ability of the particle mid-plane layer to
undergo gravitational collapse \citep{Weidenschilling1995}.

The Minimum Mass Solar Nebula has column density $\varSigma_{\rm g}=1700(r/{\rm
AU})^{-1.5}\,{\rm g\,cm^{-2}} \approx150\,{\rm g\,cm^{-2}}$ at $r=5\,{\rm AU}$
\citep{Hayashi1981}, and thus $\tilde{G}=0.042$ according to \Eq{eq:Sigma}.
Since we use $\sim$10 times higher value for $\tilde{G}$, the mean free path of
gas molecules is only $\lambda$$\sim$$10\,{\rm cm}$. Therefore our choice of
dimensionless friction times $\varOmega\tau_{\rm f}=0.25,0.50,0.75,1.00$ puts
particles in the Stokes drag force regime \citep{Weidenschilling1977a}. Here the
friction time is independent of gas density, and the Stokes number
$\varOmega\tau_{\rm f}$ is proportional to particle radius {\it squared}, so
$\varOmega\tau_{\rm f}=0.25,0.50,0.75,1.00$ correspond to physical particle
sizes ranging from $40\,{\rm cm}$ to $80\,{\rm cm}$ (see online supplement of
J07). Scaling \Eq{eq:Sigma} to more distant orbital locations gives smaller
physical particles and a gas column density closer to the Minimum Mass Solar
Nebula value, since self-gravity is more efficient in regions where the
rotational support is lower.

There are several points to be raised about our protoplanetary disc
model. The high self-gravity parameter that we use implies not only a very high
column density, but also that the gas component is close to gravitational
instability. The self-gravity parameter $\tilde{G}$ is connected to the more
commonly used $Q$ \citep[where $Q>1$ is the axisymmetric stability criterion
for a flat disc in Keplerian rotation, see][]{Safronov1960,Toomre1964} through
$\tilde{G}\approx1.6 Q^{-1}$. Thus we have $Q\approx3.2$, which means that gas
self-gravity should start to affect the dynamics (the disc is not formally
gravitationally unstable, but the disc is massive enough to slow down the
propagation of sound waves). Another issue with such a massive disc is our
assumption of ideal MHD. The high gas column density decreases the ionisation
by cosmic rays and X-rays and increases the recombination rate on dust grains
\citep{Sano+etal2000,Fromang+etal2002}.
\cite{LesurLongaretti2007,LesurLongaretti2010} have furthermore shown that the
ratio of viscosity to resistivity, the magnetic Prandtl number, affects both
small-scale and large-scale dynamics of saturated magnetorotational turbulence.
Ideally all these effects should be taken into account. However, in this paper
we choose to focus on the dynamics of solid particles in gas turbulence. Thus
we include many physical effects that are important for particles (drag forces,
self-gravity, collisions), while we ignore many other effects that would
determine the occurrence and strength of gas turbulence.  This approach allows
us to perform numerical experiments which yield insight into planetesimal
formation with relatively few free parameters.
\begin{figure}
  \includegraphics[width=\linewidth]{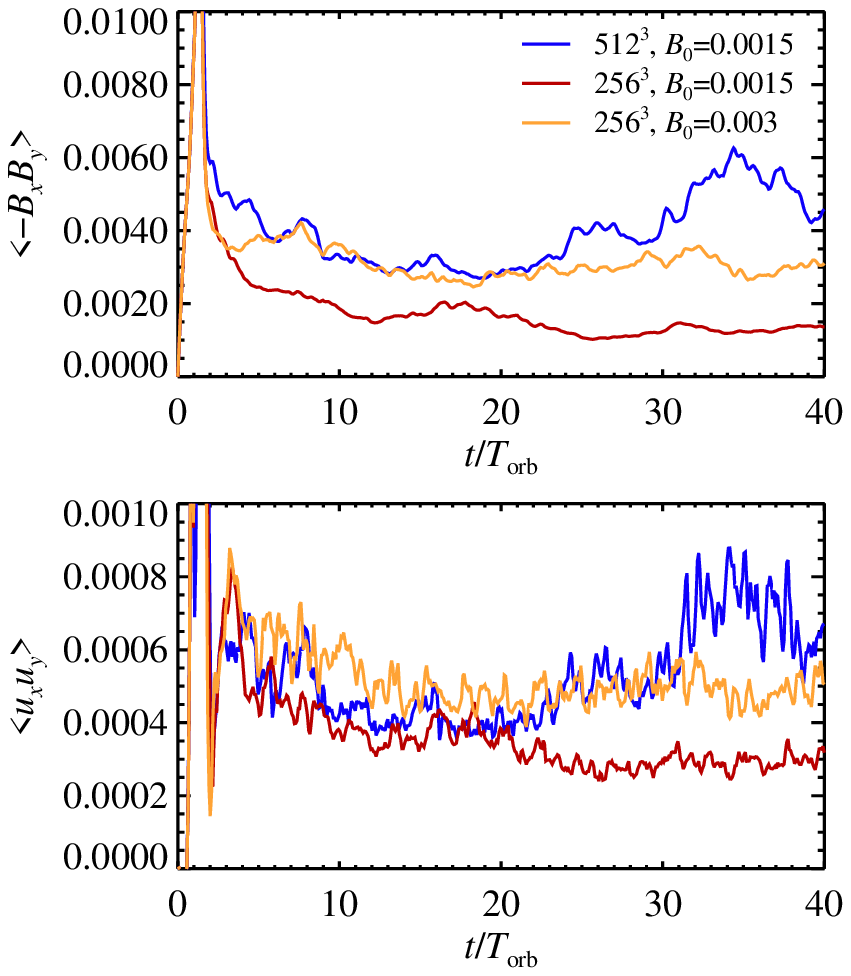}
  \caption{Maxwell and Reynolds stresses as a function of time. The Reynolds
  stress is approximately five times lower than the Maxwell stress. There is a
  marked increase in the turbulent stresses when increasing the resolution from
  $256^3$ to $512^3$ at a fixed mean vertical field $B_0=0.0015$, likely due to
  better resolution of the most unstable MRI wavelength at higher resolution.
  Using $B_0=0.003$ at $256^3$ gives turbulence properties more similar to
  $512^3$.}
  \label{f:stress_t}
\end{figure}

\subsection{Initial conditions}

The gas is initialised to have unit density everywhere in the box. The magnetic
field is constant $\vc{B}=B_0 \hat{\vc{e}}_z$. The gas velocity field is set to
be sub-Keplerian with $u_y=-\Delta v$, and we furthermore perturb all
components of the velocity field by small random fluctuations with amplitude
$\delta v=0.001$, to seed modes that are unstable to the magnetorotational
instability. In simulations with particles we give particles random initial
positions and zero velocity.
\begin{figure*}
  \begin{center}
    \includegraphics[width=0.8\linewidth]{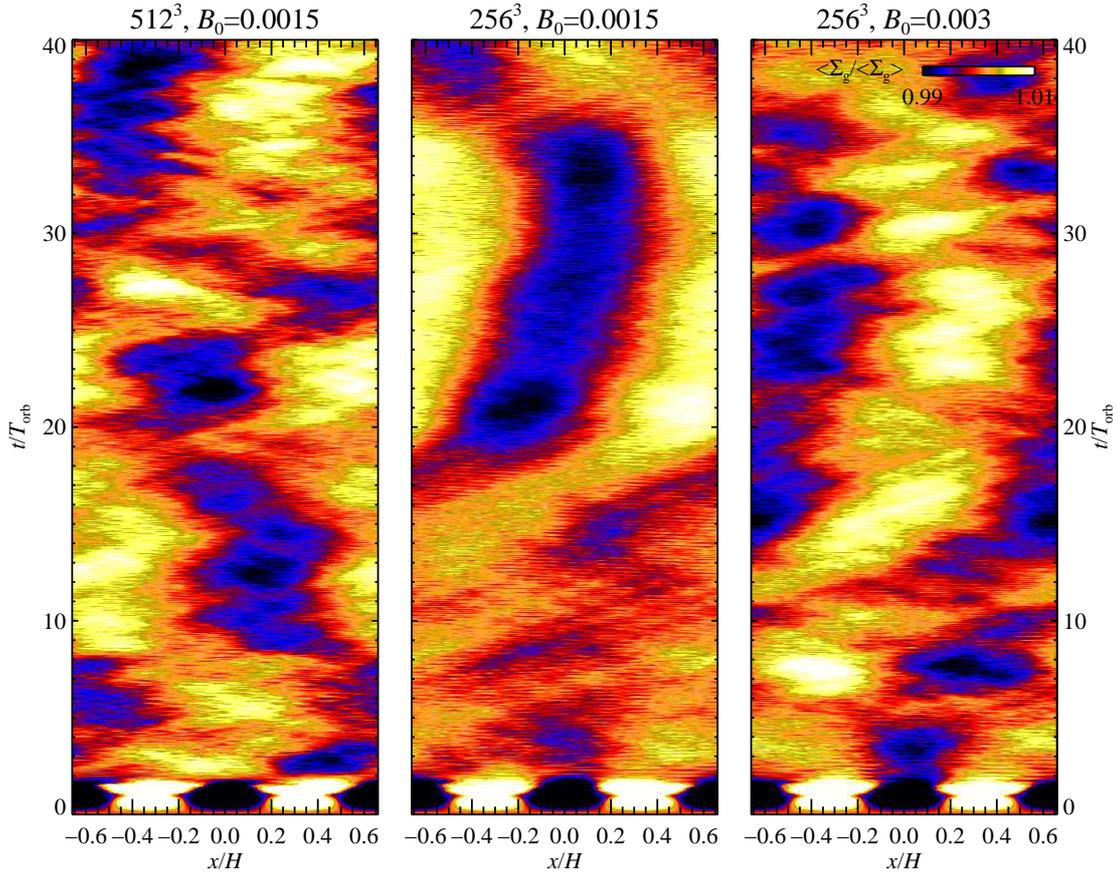}
  \end{center}
  \caption{The gas column density averaged over the azimuthal direction, as a
  function of radial coordinate $x$ and time $t$ in orbits. Large-scale
  pressure bumps appear with approximately $1\%$ amplitude at both $256^3$ and
  $512^3$ resolution.}
  \label{f:rhomx_t}
\end{figure*}

\section{Gas evolution}
\label{s:gas}

We start by describing the evolution of gas without particles, since the
large-scale geostrophic pressure bumps appearing in the gas controls particle
concentration and thus the overall ability for planetesimals to form by
self-gravity. The most important agent for driving gas dynamics is the
magnetorotational instability \citep[MRI, ][]{BalbusHawley1991} which exhibits
dynamical growth when the vertical component of the magnetic field is not too
weak or too strong. The non-stratified MRI saturates to a state of non-linear
subsonic fluctuations \citep[e.g.][]{Hawley+etal1995}. In this state there is
an outward angular momentum flux through hydrodynamical Reynolds stresses
$\langle \rho u_x u_y \rangle$ and magnetohydrodynamical Maxwell stresses
$\langle -\mu_0^{-1} B_x B_y \rangle$.

In \Fig{f:stress_t} we show the Maxwell and Reynolds stresses as a function of
time. Using a mean vertical field of $B_0=0.0015$ (corresponding to
plasma-beta of $\beta\approx9\times10^5$) the turbulent viscosity almost
triples when going from $256^3$ to $512^3$ grid points. This is in stark
contrast with zero net flux simulations that show decreasing turbulence with
increasing resolution \citep{FromangPapaloizou2007}. We interpret the behaviour
of our simulations as an effect of underresolving the most unstable wavelength
of the magnetorotational instability. Considering a vertical magnetic field of
constant strength $B_0$, the most unstable wave number of the MRI is
\citep{BalbusHawley1991} \begin{equation} k_z = \sqrt{\frac{15}{16}}
\frac{\varOmega}{v_{\rm A}} \, , \end{equation} where $v_{\rm
A}=B_0/\sqrt{\mu_0 \rho_0}$ is the Alfv\'en speed. The most unstable wavelength
is $\lambda_z=2\pi/k_z$. For $B_0=0.0015$ we get $\lambda_z \approx0.01 H$. The
resolution elements are $\delta x\approx0.005 H$ at $256^3$ and $\delta
x\approx0.0026 H$ at $512^3$. Thus we get a significant improvement in the
resolution of the most unstable wavelength when going from $256^3$ to $512^3$
grid points. Other authors \citep{Simon+etal2009,Yang+etal2009} have
reported a similar increase in turbulent activity of net-flux simulations with
increasing resolution. Our simulations show that this increase persists up to
at least $\beta\approx9\times10^5$.

The original choice of $B_0=0.0015$ was made in J07 in order to prevent the
turbulent viscosity from dropping below $\alpha=0.001$. However, we can not
obtain the same turbulent viscosity (i.e. $\alpha\sim0.001$) at higher
resolution, given the same $B_0$. For this reason we did all $256^3$
experiments on particle dynamics and self-gravity with $B_0=0.003$
($\beta\approx2\times10^5$), yielding approximately the same turbulent
viscosity as in the high-resolution simulation.

The Reynolds and Maxwell stresses can be translated into an average turbulent
viscosity \citep[following the notation of ][]{Brandenburg+etal1995},
\begin{eqnarray}
  \langle \rho u_x u_y \rangle &=& \frac{3}{2} \varOmega \nu_{\rm t}^{\rm
  (kin)} \langle \rho \rangle \, , \label{eq:nutkin} \\
  -\frac{1}{\mu_0} \langle B_x B_y \rangle &=& \frac{3}{2} \varOmega \nu_{\rm
  t}^{\rm (mag)} \langle \rho \rangle \, . \label{eq:nutmag}
\end{eqnarray}
Here $\nu_{\rm t}^{\rm (kin)}$ and $\nu_{\rm t}^{\rm (mag)}$ are the turbulent
viscosities due to the velocity field and the magnetic field, respectively. We
can further normalise the turbulent viscosities by the sound speed $c_{\rm s}$
and gas scale height $H$ \citep{ShakuraSunyaev1973},
\begin{equation}
  \alpha  = \frac{1}{c_{\rm s} H} \left[ \nu_{\rm t}^{\rm (kin)} + \nu_{\rm
  t}^{\rm (mag)} \right] \, .
\end{equation}
We thus find a turbulent viscosity of $\alpha\approx0.001$,
$\alpha\approx0.0022$, and $\alpha\approx0.003$ for runs M1, M2, and H,
respectively.

The combination of radial pressure support and two-way drag forces allows
systematic relative motion between gas and particles, which is unstable to the
streaming instability
\citep{YoudinGoodman2005,YoudinJohansen2007,JohansenYoudin2007,Miniati2010,BaiStone2010a,BaiStone2010b}.
Streaming instabilities and magnetorotational instabilities can operate in
concurrence \citep[J07,][]{Balsara+etal2009,Tilley+etal2010}. However, we find
that particles concentrate in high-pressure bumps forming due to the MRI, so
that streaming instabilities are a secondary effect in the simulations. A
necessity for the streaming instability to operate is a solids-to-gas ratio
that is locally at least unity. The particle density in the mid-plane layer is
reduced by turbulent diffusion (which is mostly caused by the MRI), so in this
way an increase in the strength of MRI turbulence can reduce the importance of
the SI. Even though streaming instabilities do not appear to be the main
driver of particle concentration in our simulations, the back-reaction drag
force of the particles on the gas can potentially play a role during the
gravitational contraction phase where local particle column densities get very
high. The high gas column density needed for gravitational collapse in the
current paper may also in reality preclude activity by the magnetorotational
instability, given the low ionisation level in the mid-plane, which would make
the streaming instability the more likely path to clumping and planetesimal
formation.

\subsection{Pressure bumps}

An important feature of magnetorotational turbulence is the emergence of
large-scale slowly overturning pressure bumps
\citep{FromangNelson2005,Johansen+etal2006}. Such pressure bumps form with a
zonal flow envelope due to random excitation of the zonal flow by large-scale
variations in the Maxwell stress \citep{Johansen+etal2009a}. Variations in the
mean field magnitude and direction has also been shown to lead to the formation
of pressure bumps in the interface region between weak and strong turbulence
\citep{Kato+etal2009,Kato+etal2010}. Pressure bumps can also be launched by a
radial variation in resistivity, e.g.\ at the edges of dead zones
\citep{Lyra+etal2008b,Dzyurkevich+etal2010}.

Large particles -- pebbles, rocks, and boulders -- are attracted to the center
of pressure bumps, because of the drag force associated with the
sub-Keplerian/super-Keplerian zonal flow envelope. In presence of a mean radial
pressure gradient the trapping zone is slightly downstream of the pressure
bump, where there is a local maximum in the combined pressure.

An efficient way to detect pressure bumps is to average the gas density field
over the azimuthal and vertical directions. In \Fig{f:rhomx_t} we show the gas
column density in the $256^3$ and $512^3$ simulations averaged over the
$y$-direction, as a function of time. Large-scale pressure bumps are clearly
visible, with spatial correlation times of approximately 10-20 orbits. The
pressure bump amplitude is around 1\%, independent of both resolution and
strength of the external field. Larger boxes have been shown to result in
higher-amplitude and longer-lived pressure bumps \citep{Johansen+etal2009a}. We
limit ourselves in this paper to a relatively small box, where we can achieve
high resolution of the gravitational collapse, but plan to model planetesimal
formation in larger boxes in the future.
\begin{figure}
  \begin{center}
    \includegraphics[width=0.8\linewidth]{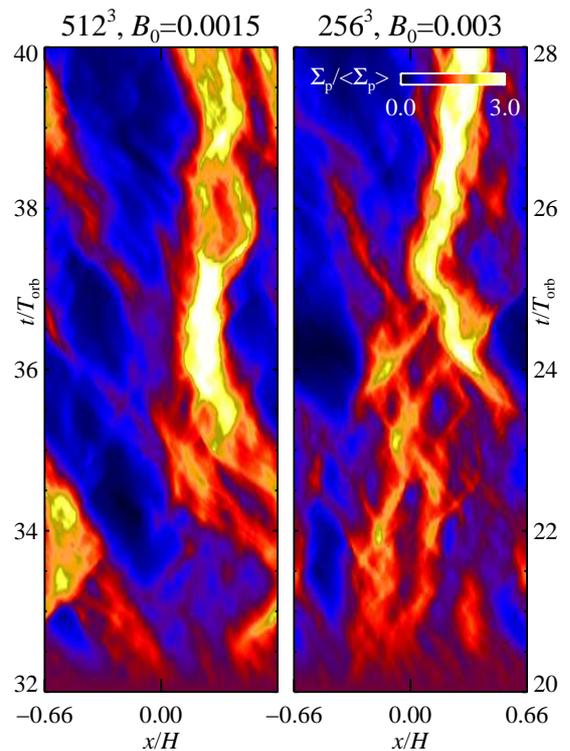}
  \end{center}
  \caption{The particle column density averaged over the azimuthal direction,
  as a function of radial coordinate $x$ and time $t$ in orbits. The starting
  time was chosen to be slightly prior to the emergence of a pressure bump
  (compare with left-most and right-most plots of \Fig{f:rhomx_t}). The
  particles concentrate slightly downstream of the gas pressure bump, with a
  maximum column density between three and four times the mean particle column
  density. The particles are between 40 and 80 cm in radius (i.e.\
  boulders) for our adopted disc model.}
  \label{f:rhopmx_t}
\end{figure}

\section{Particle evolution}
\label{s:particles}

We release the particles at a time when the turbulence has saturated, but
choose a time when there is no significant large-scale pressure bump present.
Thus we choose $t=20T_{\rm orb}$ for the $256^3$ simulation and $t=32T_{\rm
orb}$ for the $512^3$ simulation (see left-most and right-most plot of
\Fig{f:rhomx_t}). In the particle simulations we always use a mean vertical
field $B_0=0.003$ at $256^3$ to get a turbulent viscosity more similar to
$512^3$. The four friction time bins ($\varOmega\tau_{\rm
f}=0.25$,$0.50$,$0.75$,$1.00$) correspond to particle sizes between 40 and 80
cm.
\begin{figure*}[t!]
  \begin{center}
    \includegraphics[width=0.39\linewidth]{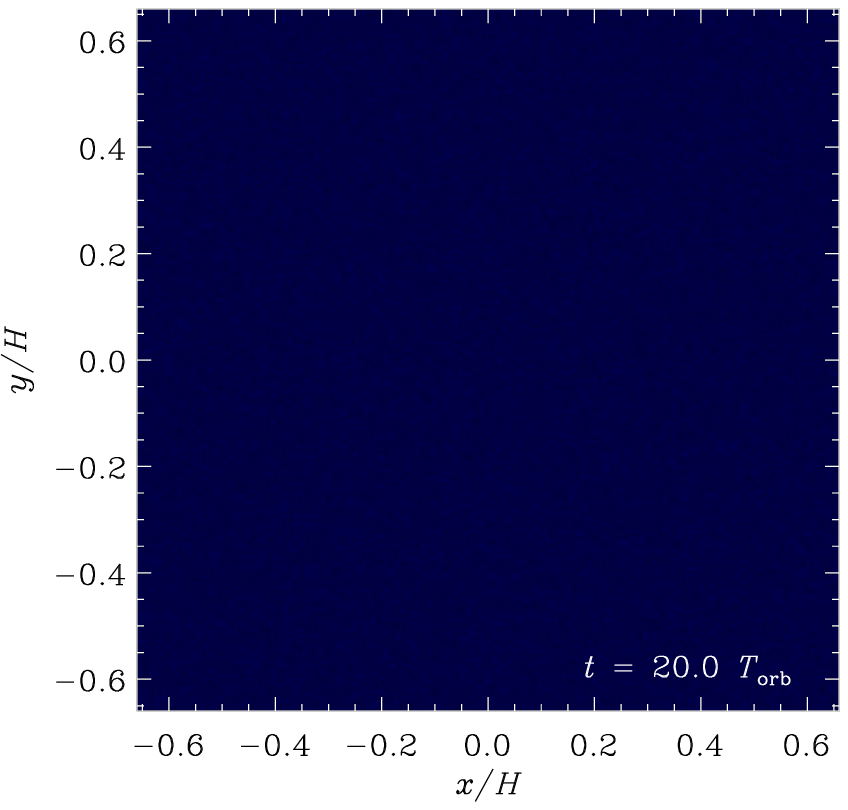}
    \includegraphics[width=0.39\linewidth]{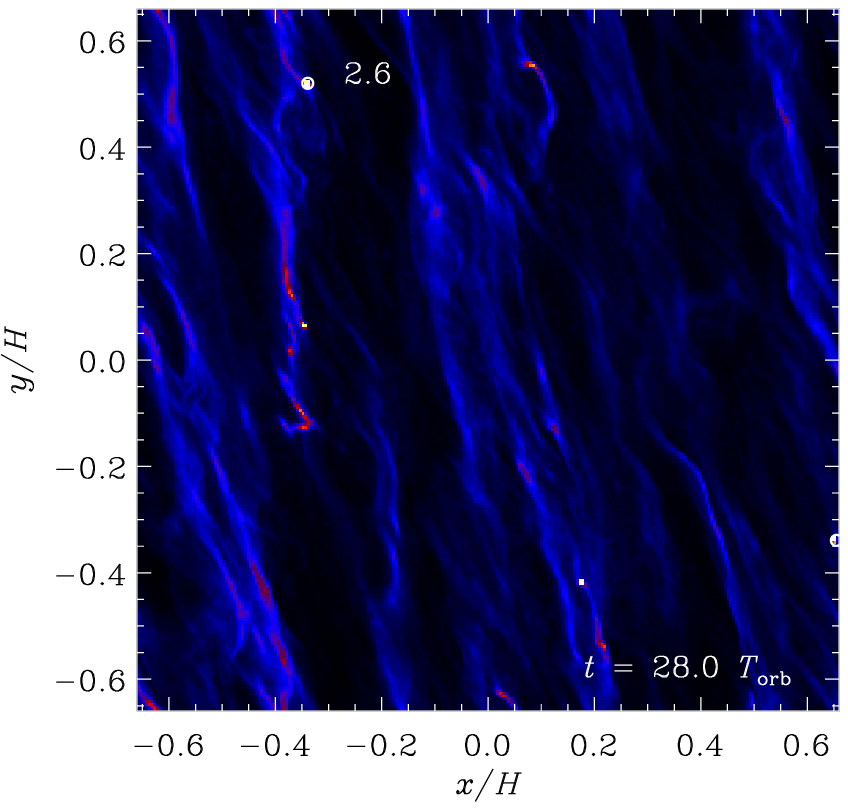}
    \includegraphics[width=0.39\linewidth]{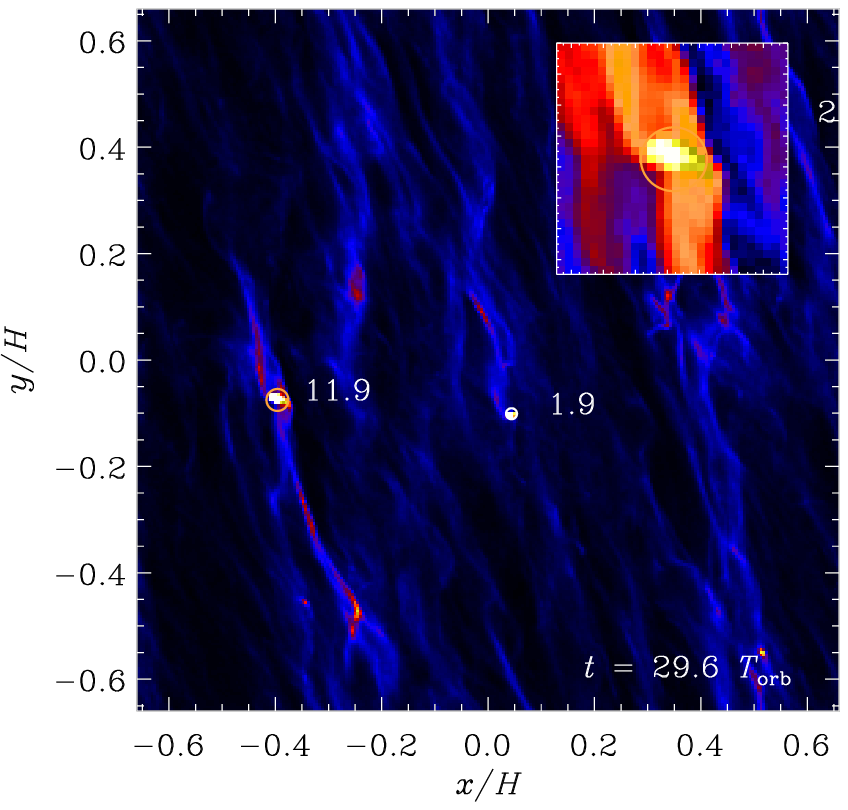}
    \includegraphics[width=0.39\linewidth]{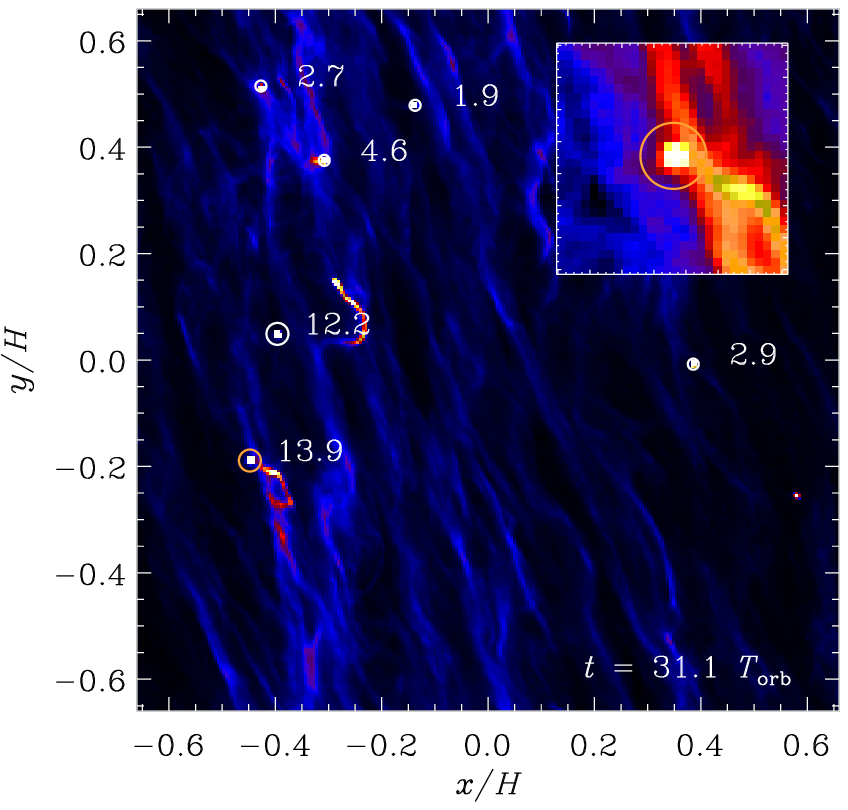}
    \includegraphics[width=0.39\linewidth]{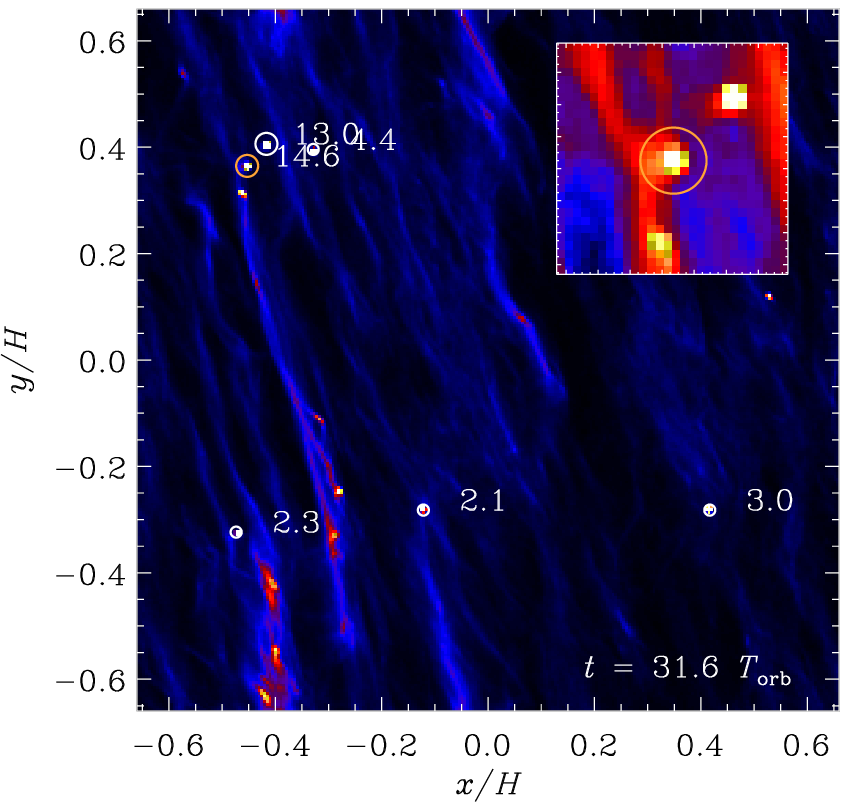}
    \includegraphics[width=0.39\linewidth]{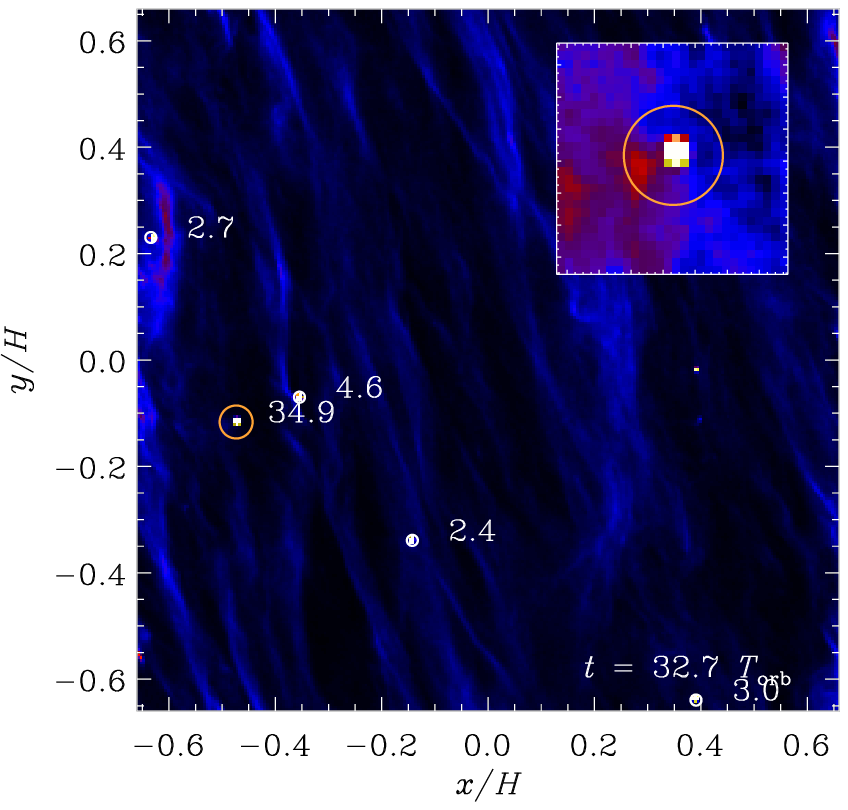}
  \end{center}
  \caption{The particle column density as a function of time after self-gravity
  is turned on at $t=20.0T_{\rm orb}$, for run M2 ($256^3$ grid cells with
  $8\times10^6$ particles). Each gravitationally bound clump is labelled by its
  Hill mass in units of Ceres masses. The insert shows an enlargement of the
  region around the most massive bound clump. The most massive clump at the end
  of the simulation contains a total particle mass of $34.9$ Ceres masses,
  partially as the result of a collision between a $13.0$ and a $14.6$ Ceres
  mass clump occurring at a time around $t=31.6 T_{\rm orb}$.}
  \label{f:selfg256}
\end{figure*}
\begin{figure*}[t!]
  \begin{center}
    \includegraphics[width=0.39\linewidth]{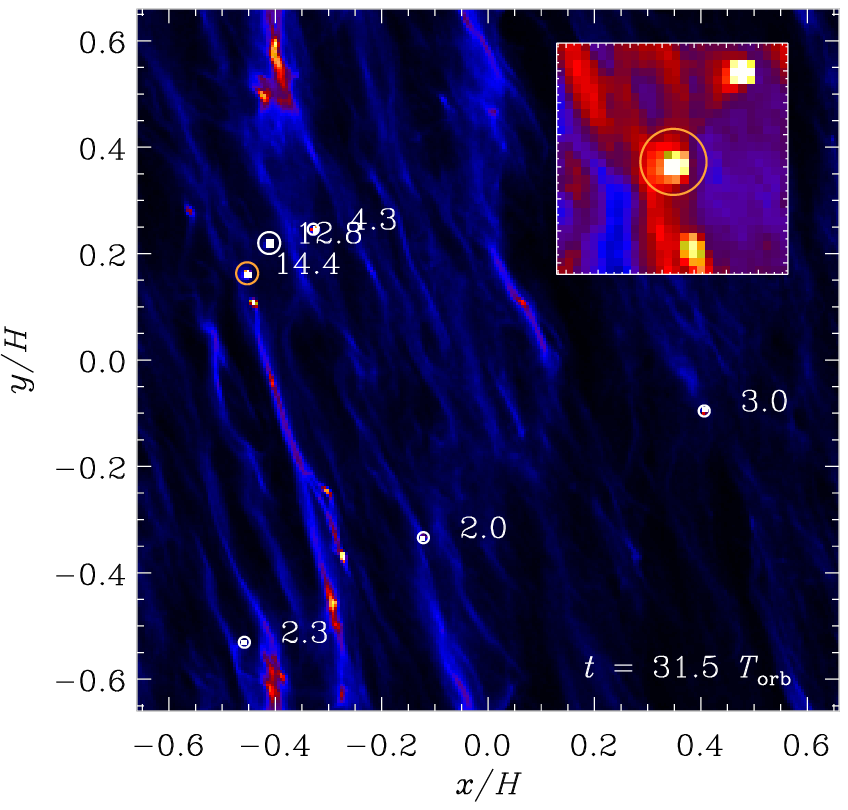}
    \includegraphics[width=0.39\linewidth]{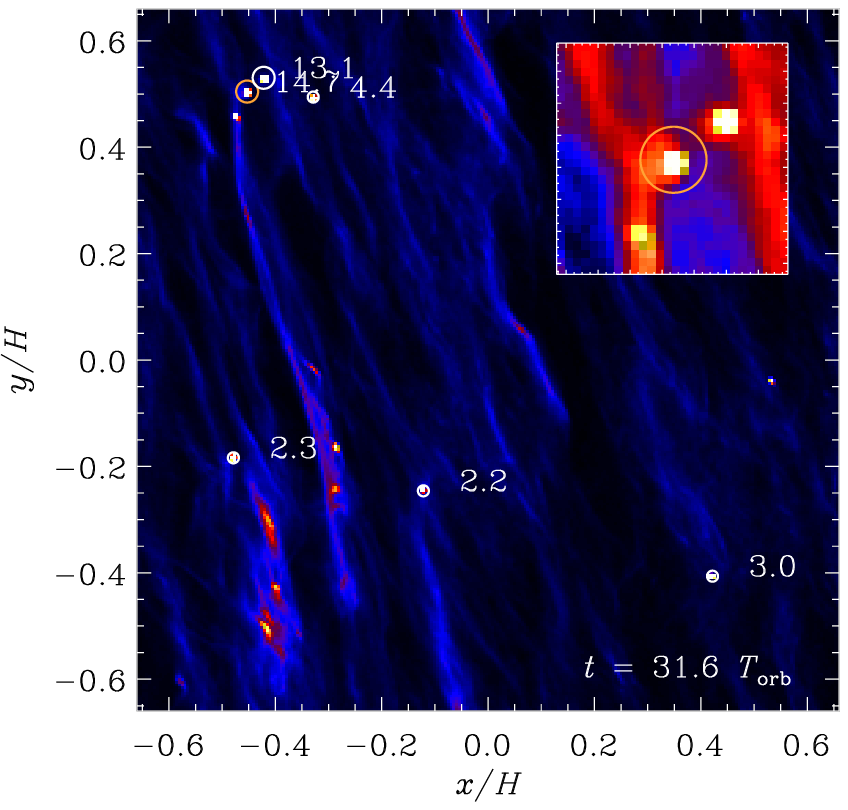}
    \includegraphics[width=0.39\linewidth]{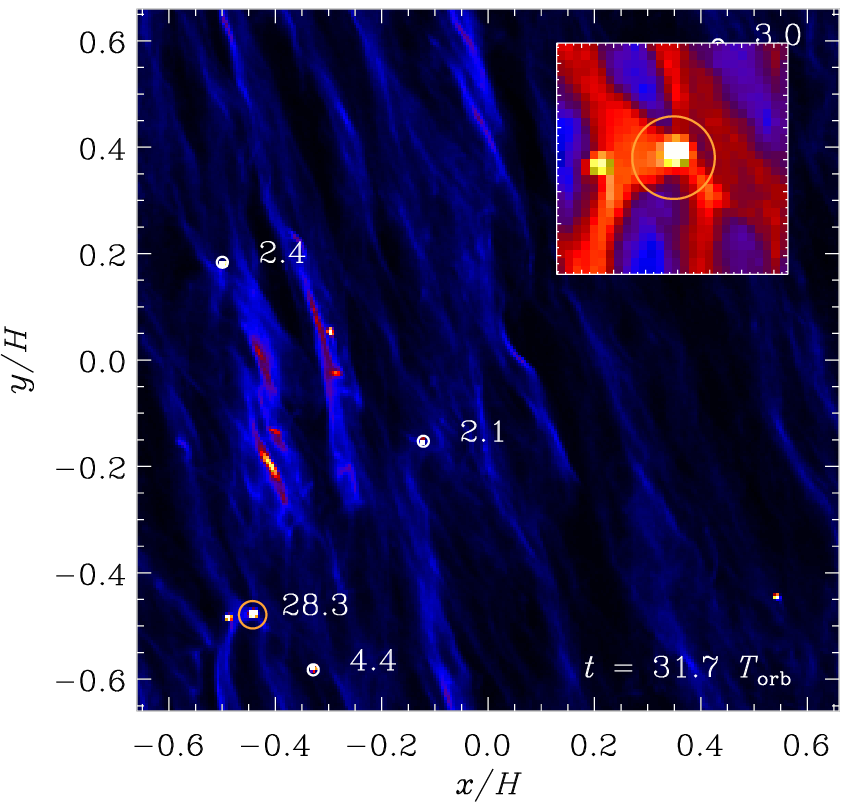}
    \includegraphics[width=0.39\linewidth]{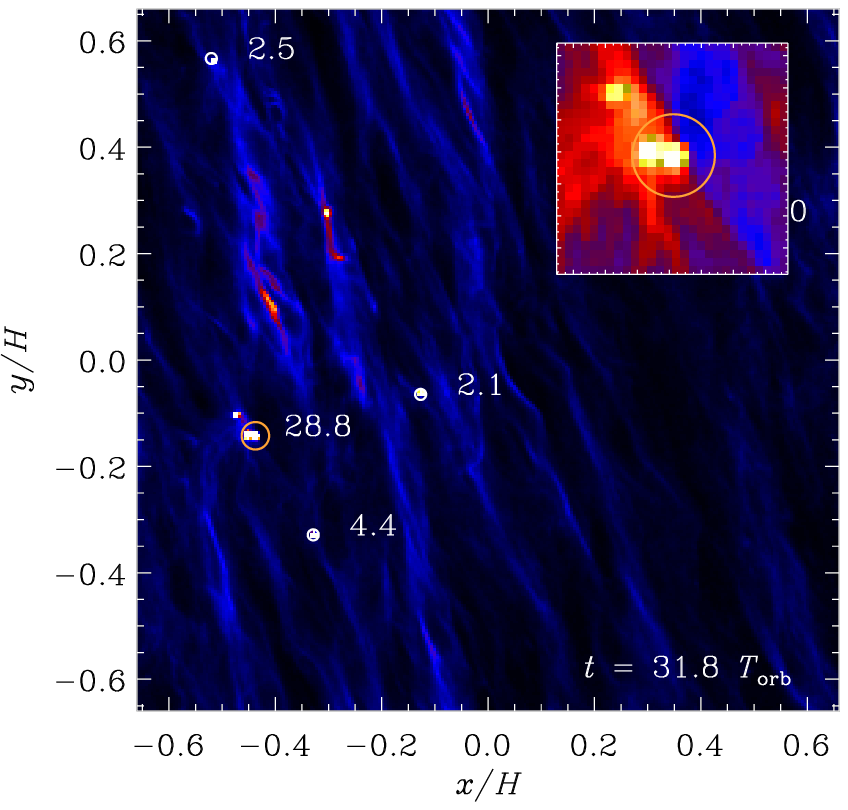}
    \includegraphics[width=0.39\linewidth]{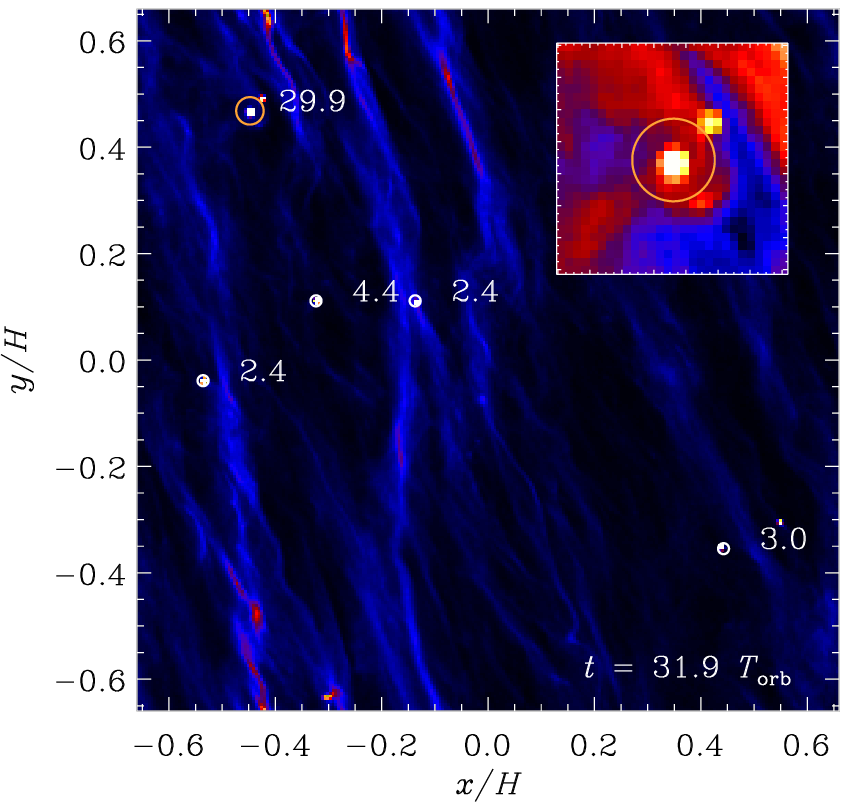}
    \includegraphics[width=0.39\linewidth]{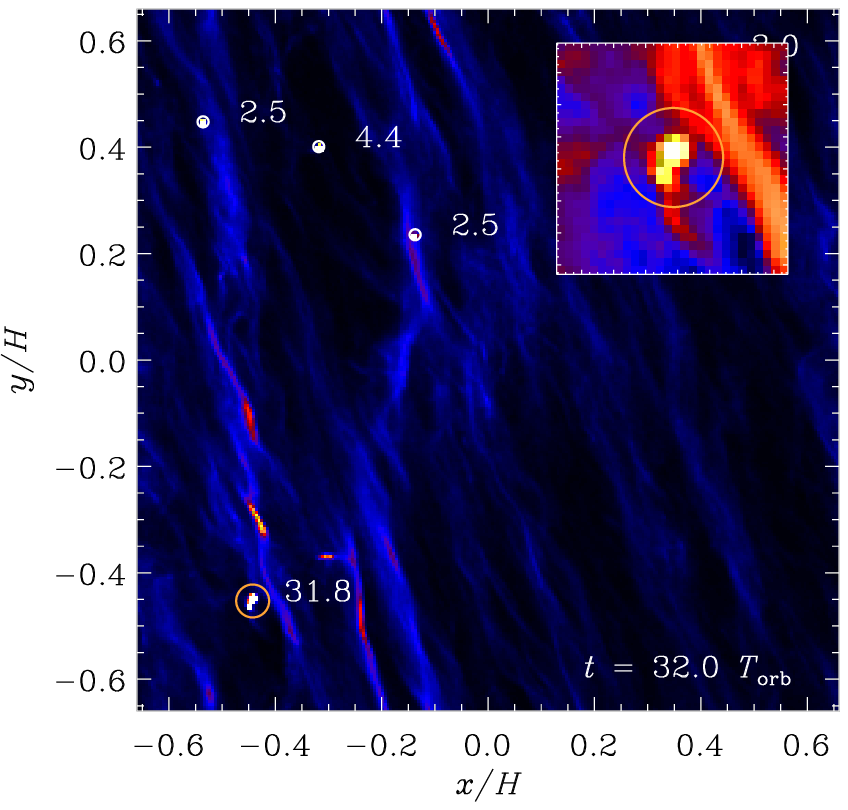}
  \end{center}
  \caption{Temporal zoom in on the collision between three clumps
  (planetesimals) in the moderate resolution run M2. Two clumps with a radial
  separation approximately $0.05 H$ shear past each other, bringing their Hill
  spheres in contact (first two panels). The clumps first pass through each
  other (panels three and four), but eventually merge (fifth panel). Finally a
  much lighter clump collides with the newly formed merger product (sixth
  panel).}
  \label{f:merge}
\end{figure*}

The particles immediately fall towards the mid-plane of the disc, before
finding a balance between sedimentation and turbulent stirring.
\Fig{f:rhopmx_t} shows how the presence of gas pressure bumps has a dramatic
influence on particle dynamics. The particles display column density
concentrations of up to 4 times the average density just downstream of the
pressure bumps. At this point the gas moves close to Keplerian, because the
(positive) pressure gradient of the bump balances the (negative) radial
pressure gradient there. The column density concentration is relatively
independent of the resolution, as expected since the pressure bump amplitude is
almost the same.

\section{Self-gravity -- moderate resolution}
\label{s:sgmod}

In the moderate-resolution simulation ($256^3$) we release particles and start
self-gravity simultaneously at $t=20T_{\rm orb}$. This is different from the
approach taken in J07 where self-gravity was turned on after the particles had
concentrated in a pressure bump. Thus we address concerns that the
continuous self-gravity interaction of the particles would stir up the particle
component and prevent the gravitational collapse. After releasing particles we
continue the simulation for another thirteen orbits. Some representative
particle column density snapshots are shown in \Fig{f:selfg256}. As time
progresses the particle column density increases in high-pressure structures
with power on length scales ranging from a few percent of a scale height to the
full radial domain size. Self-gravity becomes important in these overdense
regions, so some local regions begin to contract radially under their own
weight, eventually reaching the Roche density and commencing a fully 2-D
collapse into discrete clumps.

The Hill sphere of each bound clump is indicated in \Fig{f:selfg256}, together
with the mass of particles encompassed inside the Hill radius (in units of the
mass of the dwarf planet Ceres). We calculate the Hill radius of clumps at a
given time by searching for the point of highest particle column density in the
$x$-$y$ plane. We first consider a ``circle'' of radius one grid point and
calculate the two terms relevant for determination of the Hill radius -- the
tidal term $3\varOmega^2 R$ and the gravitational acceleration $G M_{\rm
p}/R^2$ of a test particle at the edge of the Hill sphere due to the combined
gravity of particles inside the Hill sphere. The mass $M_{\rm p}$ contained in
a cylinder of radius $R$ must fulfil
\begin{equation}
  M_{\rm p} \ge \frac{3 \varOmega^2 R^3}{G} \, .
\end{equation}
The natural constant $G$ is set by the non-dimensional form of the Poisson
equation,
\begin{equation}
  \nabla^2_H \frac{\varPhi}{c_{\rm s}^2} = \frac{4 \pi G}{\varOmega^2/\rho_0}
  \frac{\rho_{\rm p}}{\rho_0} \, .
\end{equation}
Here $\nabla^2_H \equiv \dpa^2/\dpa(x/H)^2 + \dpa^2/\dpa(y/H)^2 +
\dpa^2/\dpa(z/H)^2$. Using natural units for the simulation, with $c_{\rm
s}=\varOmega=H=\rho_0=1$, together with our choice of
\begin{equation}
  \tilde{G} = \frac{4 \pi G}{\varOmega^2/\rho_0} = 0.5 \, ,
\end{equation}
we obtain an expression for the gravitational constant $G$. We then check the
validity of the expression
\begin{equation}
  M_{\rm p} = \sum_{i,j} n_z \overline{\rho}_{ij} \delta V \ge \frac{12 \pi
  \rho_0 R^3}{\tilde{G}} \, ,
  \label{eq:hillcrit}
\end{equation}
where $\overline{\rho}_{ij}=n_z^{-1} \sum_k \rho_{ijk}$ is the vertically
averaged mass density at grid point $(i,j)$ and $\delta V$ is the volume of a
grid cell. It is convenient to work with $\overline{\rho}_{ij}$ since this
vertical average has been output regularly by the code during the simulation.
The sum in \Eq{eq:hillcrit} is taken over all grid points lying within the
circle of radius $R$ centred on the densest point. We continue to increase $R$
in units of $\delta x$ until the boundness criterion is no longer fulfilled.
This defines the Hill radius $R$. Strictly speaking our method for finding the
Hill radius is only valid if the particles were distributed in a spherically
symmetric way. In reality particles are spread across the mid-plane with a
scale height of approximately $0.04 H$. We nevertheless find by inspection
that the calculated Hill radii correspond well to the regions where the
particle flow appears dominated by the self-gravity rather than the Keplerian
shear of the main flow and that the mass within the Hill sphere does not
fluctuate because of the inclusion of non-bound particles.
\begin{figure*}
  \begin{center}
    \includegraphics[width=7.5cm]{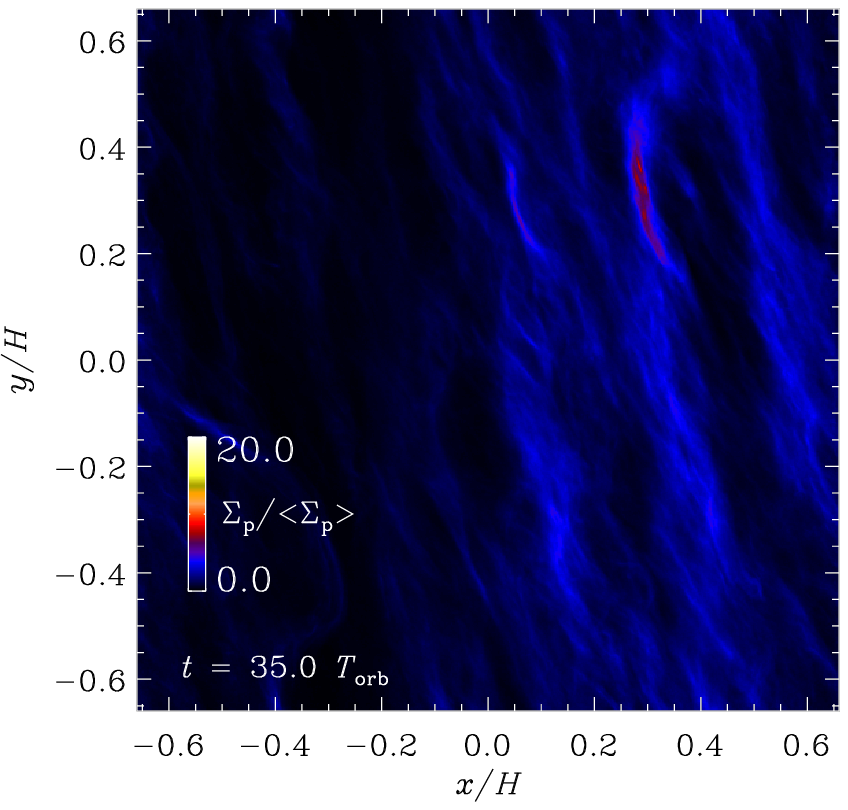}
    \includegraphics[width=7.5cm]{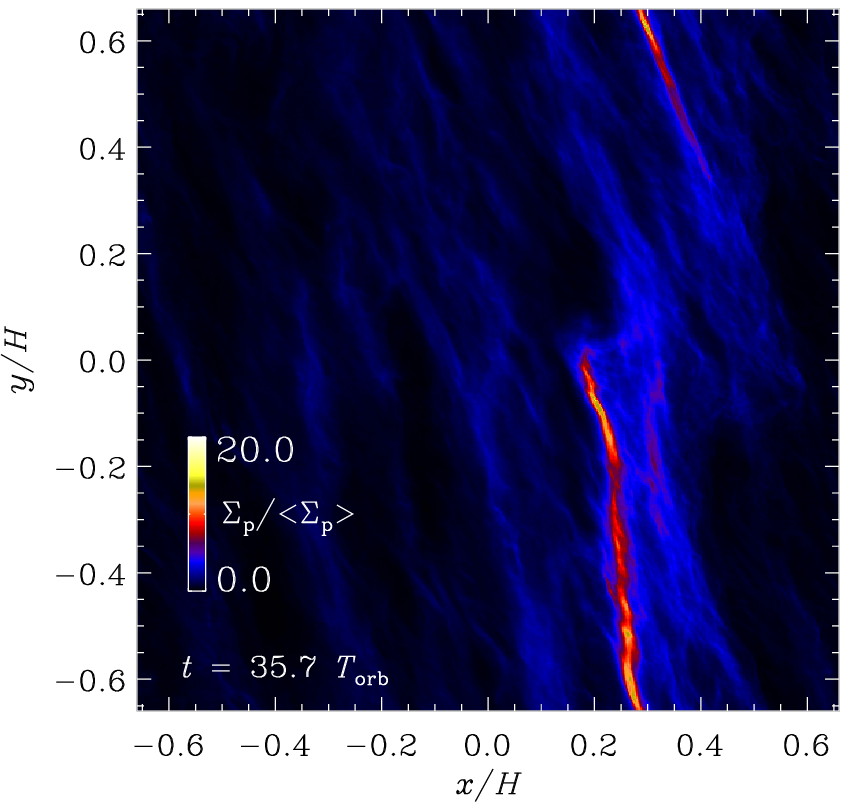}
    \includegraphics[width=7.5cm]{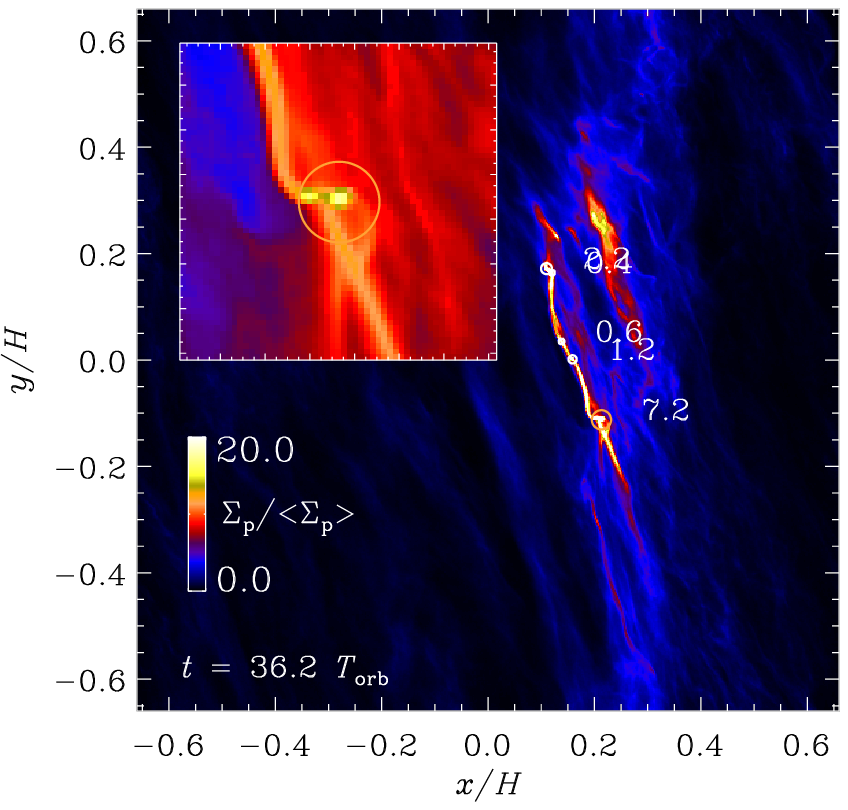}
    \includegraphics[width=7.5cm]{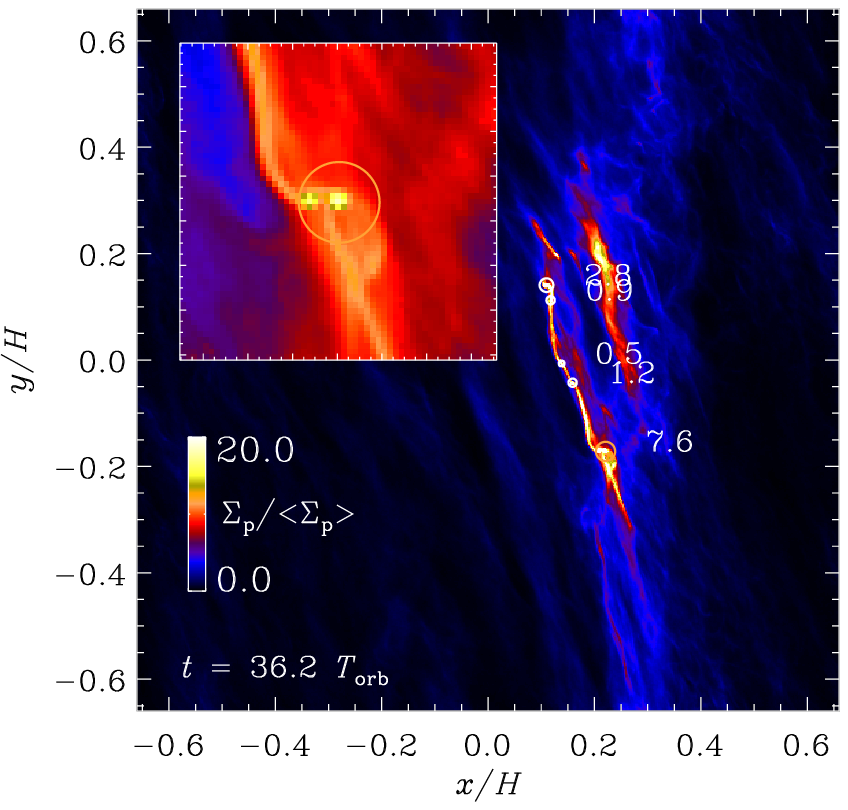}
    \includegraphics[width=7.5cm]{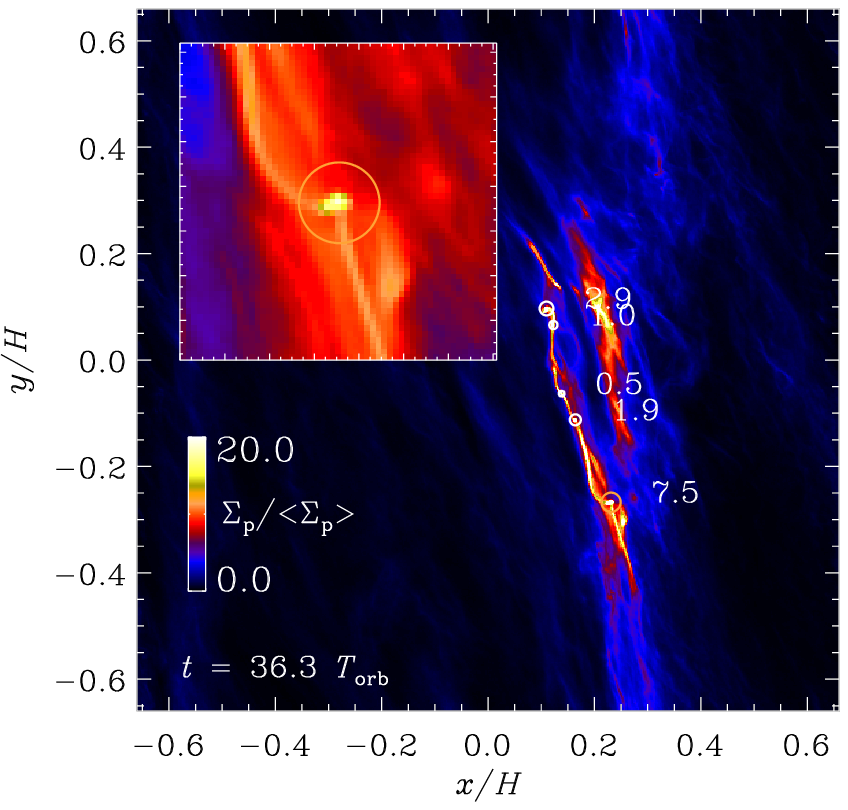}
    \includegraphics[width=7.5cm]{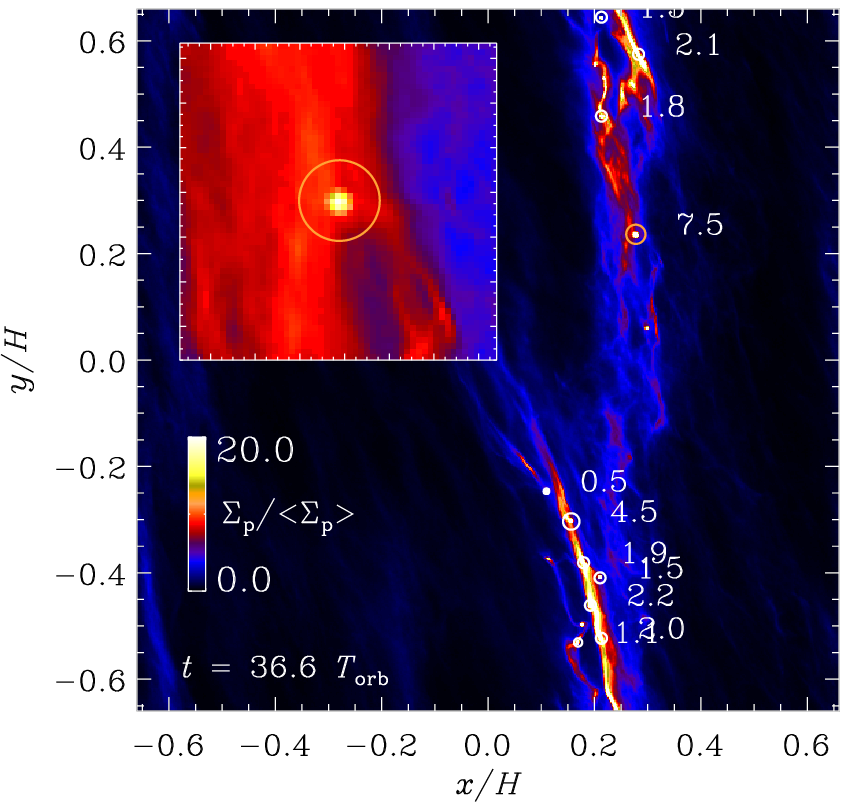}
  \end{center}
  \caption{The particle column density as a function of time after self-gravity
  is turned on after $t=35.0T_{\rm orb}$ in the high-resolution simulation (run
  H with $512^3$ grid cells and $64\times10^6$ particles). Two clumps condense
  out within each other's Hill spheres and quickly merge. At the end of the
  simulation bound clumps have masses between $0.5$ and $7.5$ $M_{\rm Ceres}$.}
  \label{f:selfg512}
\end{figure*}

The particle-mesh Poisson solver based on Fast Fourier Transforms can not
consider the gravitational potential due to structure within a grid cell. From
the perspective of the gravity solver the smallest radius of a bound object is
thus the grid cell length $\delta x$. This puts a lower limit to the mass of
bound structures, since the Hill radius can not be smaller than a grid cell,
\begin{equation}
  R_{\rm H} = \left( \frac{G M_{\rm min}}{3\varOmega^2} \right)^{1/3} \approx
  \delta x \, .
\end{equation}
This gives a minimum mass of
\begin{equation}
  \frac{M_{\rm min}}{M_\star} = 3 \left( \frac{\delta x}{r} \right)^3 = 3
  \left( \frac{H}{r} \right)^3 \left( \frac{\delta x}{H} \right)^3 \, .
\end{equation}
Using $M_\star=2.0\times10^{33}\,{\rm g}$, $H/r=0.05$ and $\delta x=0.0052 H$
($256^3$) or $\delta x=0.0026$ ($512^3$), we get the minimum mass of $M_{\rm
min}\approx0.11 M_{\rm Ceres}$ at $256^3$ and $M_{\rm min}\approx0.014 M_{\rm
Ceres}$ at $512^3$. Less massive objects will inevitably be sheared out due to
the gravity softening.

\Fig{f:selfg256} shows that a number of discrete particle clumps condense out
of the turbulent flow during the 13 orbits that are run with self-gravity. The
most massive clump has the mass of approximately 35 Ceres masses at the end of
the simulation, while four smaller clumps have masses between $2.4$ and $4.6$
Ceres masses. The smallest clumps are more than ten times more massive than the
minimum resolvable mass.

\subsection{Planetesimal collision}
\label{s:collision}

The 35 Ceres masses particle clump visible in the last panel of
\Fig{f:selfg256} is partially the result of a collision between a $13.0$ and a
$14.6$ Ceres mass clump at a time around $t=31.6 T_{\rm orb}$. The
collision is shown in greater detail in \Fig{f:merge}. The merging starts when
two clumps with radial separation of approximately $0.05 H$ shear past
each other, bringing their Hill spheres in contact. The less massive
clumps passes first directly through the massive clump, appearing
distorted on the other side, before merging completely. A third clump
collides with the collision product shortly afterwards, adding another 3.5
Ceres masses to the clump.
\begin{figure*}
  \begin{center}
    \includegraphics[width=0.8\linewidth]{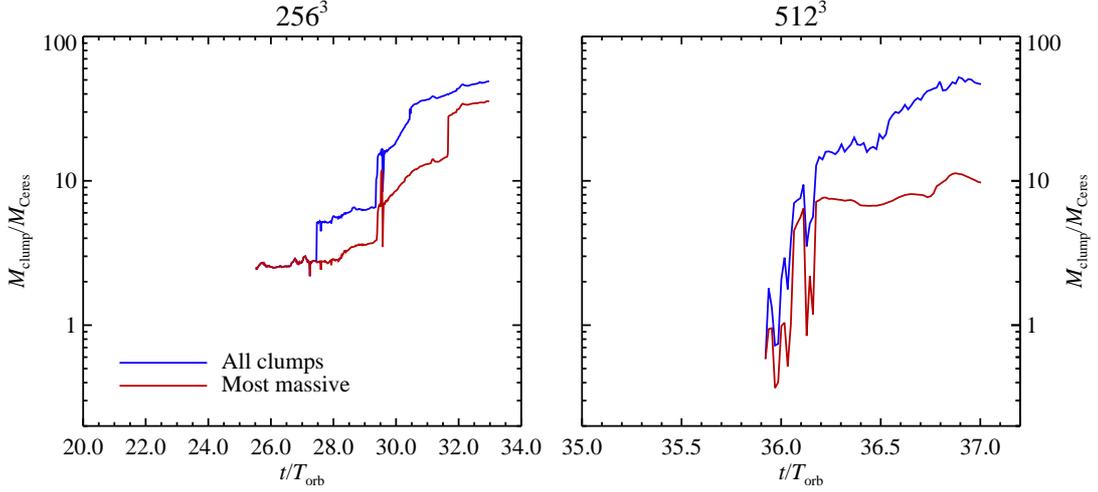}
  \end{center}
  \caption{The total bound mass and the mass in the most massive
  gravitationally bound clump as a function of time. The left panel shows the
  result of the moderate-resolution simulation (M2). Around a time of $30
  T_{\rm orb}$ there is a condensation event that transfers many particles to
  bound clumps. Two orbits later, at $32 T_{\rm orb}$, the two most massive
  clumps collide and merge. The right panel shows the high-resolution
  simulation (H). The total amount of condensed material is comparable to M2,
  but the mass of the most massive clump is smaller. This may be a result
  either of increased resolution in the self-gravity solver or of the limited
  time span of the high-resolution simulation. The total particle mass for both
  resolutions is $M_{\rm tot}\approx460\,M_{\rm Ceres}$. Only around 10\% of
  the mass is converted into planetesimals during the time-span of the
  simulations.}
  \label{f:mclump_t}
\end{figure*}
\begin{figure*}
  \begin{center}
    \includegraphics[width=0.8\linewidth]{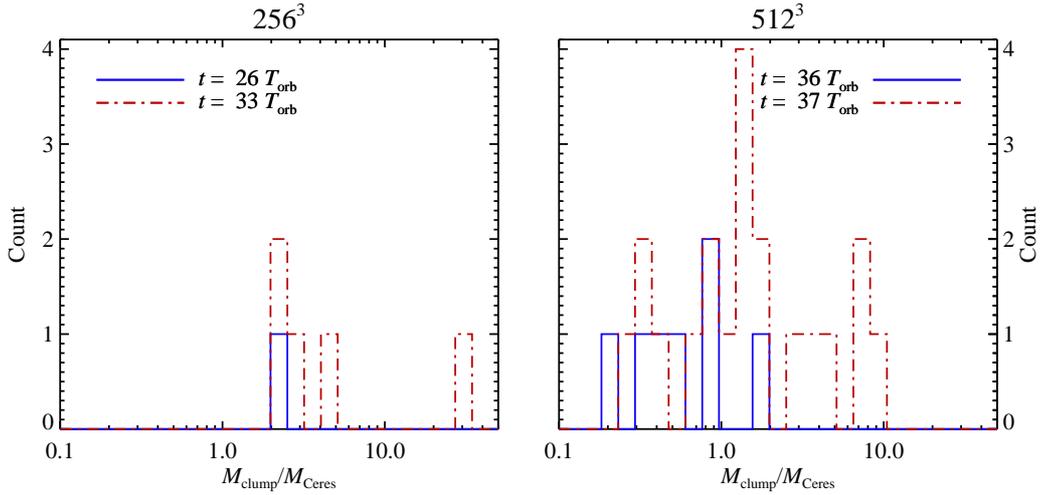}
  \end{center}
  \caption{Histogram of clump masses after first production of bound
  clumps and at the end of the simulation. At moderate resolution (left panel)
  only a single clump condenses out initially, but seven orbits later there are
  five clumps, including the 30+ $M_{\rm Ceres}$ object formed by merging. At
  high resolution (right panel) the initial planetesimal burst leads to the
  formation of many sub-Ceres-mass clumps. The most massive clump is
  similar to what forms initially in the moderate-resolution run.}
  \label{f:mclump_histo}
\end{figure*}

The particle-mesh self-gravity solver does not resolve particle self-gravity on
scales smaller than a grid cell. The bound particle clumps therefore stop their
contraction when the size of a grid cell is reached. This exaggerated size
increases the collisional cross section of planetesimal artificially. The
clumps behave aerodynamically like a collection of dm-sized particles, while a
single dwarf planet sized would have a much longer friction time. Therefore the
planetesimal collisions that we observe are not conclusive evidence of a
collisionally evolving size distribution. Future convergence tests at extremely
high resolution ($1024^3$ or higher), or mesh refinement around the clumps,
will be needed to test the validity of the planetesimal mergers.

The system is however {\it not} completely dominated by the discrete gravity
sources. A significant population of free particles are present even after
several bound clumps have formed. Those free particles can act like dynamical
friction and thus allow close encounters to lead to merging or binary formation
\citep{Goldreich+etal2002}. In the high-resolution simulation presented in the
next section we find clear evidence of a trailing spiral density structure that
is involved in the collision between two planetesimals.

\section{Self-gravity -- high resolution}
\label{s:sghigh}

In Section \ref{s:particles} we showed that particle concentration is
maintained when going from $256^3$ to $512^3$ grid cells. In this section we
show that the inclusion of self-gravity at high resolution leads to rapid
formation of bound clumps similar to what we observe at moderate resolution.
Given the relatively high computational cost of self-gravity simulations we
start the self-gravity at $t=35T_{\rm orb}$ in the $512^3$ simulation, three
orbits after inserting the particles. The evolution of particle column density
is shown in \Fig{f:selfg512}. Due to the smaller grid size bound particle
clumps appear visually smaller than in the $256^3$ simulation. The increased
compactness of the planetesimals can potentially decrease the probability for
planetesimal collisions (Sect.\ \ref{s:collision}), which makes it important to
do convergence tests.

The high-resolution simulation proceeds much as the moderate-resolution
simulation. Panels 1 and 2 of \Fig{f:selfg512} show how overdense bands of
particles contract radially under their own gravity. The increased resolution
of the self-gravity solver allows for a number of smaller planetesimals to
condense out as the bands reach the local Roche density at smaller radial
scales (panel 3). Two of the clumps are born within each other's Hill spheres.
They merge shortly after into a single clump (panel 5). This clump has grown to
$7.5$ $M_{\rm Ceres}$ at the end of the simulation, which is the most massive
clump in a population of clumps with masses between $0.5$ and $7.5$ $M_{\rm
Ceres}$.

Although we do not reach the same time span as in the low resolution
simulation, we do observe two bound clumps colliding. However, the situation is
different, since the colliding clumps form very close to each other and merge
almost immediately. An interesting aspect is the presence of a particle density
structure trailing the less massive clump. The gravitational torque from this
structure can play an important role in the collision between the two clumps,
since the clumps initially appear to orbit each other progradely. This
confirms the trend observed in \cite{JohansenLacerda2010} for particles to be
accreted with prograde angular momentum in the presence of drag forces, which
can explain why the largest asteroids tend to spin in the prograde
direction\footnote{Prograde rotation is not readily acquired in standard
numerical models where planetesimals accumulate in a gas-free environment,
although planetary birth in rotating self-gravitating gas blobs was recently
been put forward to explain the prograde rotation of
planets \citep{Nayakshin2011}.}. The gravitational torque from the trailing
density structure would be negative in that case and cause the relative
planetesimal orbit to shrink.

\subsection{Clump masses}

In \Fig{f:mclump_t} we show the total mass in bound clumps as a function of
time. Finding the physical mass of a clump requires knowledge of the scale
height $H$ of the gas, as that sets the physical unit of length. The
self-gravity solver in itself only needs to know $\tilde{G}=4 \pi
G/(\varOmega^2/\rho_0)$, which is effectively a combination of density and
length scale. When quoting the physical mass we assume orbital distance
$r=5\,{\rm AU}$ and aspect ratio $H/r=0.05$. The total mass in particles in the
box is $M_{\rm tot}=0.01\varSigma_{\rm gas}L_x L_y \approx 460\,M_{\rm Ceres}$,
with $\tilde{G}=0.5$ and $\varSigma_{\rm gas}=1800\,{\rm g\,cm^{-2}}$ from
\Eq{eq:Sigma}.

In both simulations approximately 50 $M_{\rm Ceres}$ of particles are present
in bound clumps at the end of the simulation. However, self-gravity was turned
on and sustained for very different times in the two different simulations. In
the moderate-resolution simulation most mass is bound in a single clump at the
end of the simulation. The merger event discussed in Sect.\ \ref{s:collision}
is clearly visible around $t=32 T_{\rm orb}$.

\Fig{f:mclump_histo} shows histograms of the clump masses for moderate
resolution (left panel) and high resolution (right panel). At moderate
resolution initially only a single clump forms, but seven orbits later there
are 5 bound clumps, all of several Ceres masses. The high-resolution run
produces many more small clumps in the initial planetesimal formation burst.
This is likely an effect of the ``hot start'' initial condition where we turn
on gravity during a concentration event as the high particle density allows
smaller regions to undergo collapse.

\section{Summary and discussion}
\label{s:discussion}

In this paper we present (a) the first $512^3$ grid cell simulation of dust
dynamics in turbulence driven by the magnetorotational instability and (b) a
long time integration of the system performed at $256^3$ grid cells. Perhaps
the most important finding is that large-scale pressure bumps and zonal flows
in the gas appear relatively independent of the resolution. The same is true
for particle concentration in these pressure bumps. While saturation properties
of MRI turbulence depend on the implicit or explicit viscosity and resistivity
\citep{LesurLongaretti2007,FromangPapaloizou2007,Davis+etal2010}, the emergence
of large-scale zonal flows appears relatively independent of resolution
\citep[this work,][]{Johansen+etal2009a} and numerical scheme
\citep{FromangStone2009,StoneGardiner2010}. Particle concentration in pressure
bumps can have profound influence on particle coagulation by supplying an
environment with high densities and low radial drift speeds
\citep{Brauer+etal2008b}, and on formation of planetesimals by gravitational
contraction and collapse of overdense regions (this work, J07).

A direct comparison between the moderate-resolution and the high-resolution
simulation is more difficult after self-gravity is turned on. The appearance of
gravitationally bound clumps is inherently stochastic, as is the amplitude and
phase of the first pressure bump to appear. The comparison is furthermore
complicated by the extreme computational cost of the high-resolution
simulation, which allowed us to evolve the system only for a few orbits after
self-gravity is turned on.

A significant improvement over J07 is that the moderate-resolution simulation
could be run for much longer time. Therefore we were able to start self-gravity
shortly after MRI turbulence had saturated, and to follow the system for more
than ten orbits. In J07 self-gravity was turned on during a concentration
event, precluding the concurrent evolution of self-gravity and concentration.
This ``hot start'' may artificially increase the number of the forming
planetesimals. The ``cold start'' initial conditions presented here lead to a
more gradual formation of planetesimals over more than ten orbits. Still the
most massive bound clump had grown to approximately 35 $M_{\rm Ceres}$ at the
end of the simulation and was still gradually growing.

The high-resolution simulation was given a ``hot start'', to focus
computational resources on the most interesting part of the evolution. As
expected these initial conditions allow a much higher number of smaller
planetesimals to condense out of the flow. The most massive planetesimal at the
end of the high-resolution simulation contained 7.5 $M_{\rm Ceres}$ of
particles, but this ``planetesimal'' is accompanied by a number of bound clumps
with masses from $0.5$ to $4.5$ $M_{\rm Ceres}$.

The first clumps to condense out is on the order of a few Ceres masses in
both the moderate-resolution simulation and the high-resolution simulation. The
high-resolution simulation produced additionally a number of sub-Ceres-mass
clumps. It thus appears that the higher-resolution simulation samples the
initial clump function down to smaller masses. This observation strongly
supports the use of simulations of even higher resolution in order to study the
broader spectrum of clump masses. Higher resolution will also allow studying
simultaneously the concentration of smaller particles in smaller eddies and
their role in the planetesimal formation process \citep{Cuzzi+etal2010}.

We emphasize here the difference between the Initial Mass Function of
gravitationally bound clumps and of planetesimals. The first can be studied in
the simulations that we present in this paper, while the actual masses and
radii of planetesimal that form in the bound clumps will require the inclusion
of particle shattering and particle sticking. \cite{ZsomDullemond2008} and
\cite{Zsom+etal2010} used a representative particle approach to model
interaction between superparticles in a 0-D particle-in-box approach, based on
a compilation of laboratory results for the outcome of collisions depending on
particle sizes, composition and relative speed \citep{Guettler+etal2010}. We
plan to implement this particle interaction scheme in the Pencil Code and
perform simulations that include the concurrent action of hydrodynamical
clumping and particle interaction. This approach will ultimately be needed to
determine whether each clump forms a single massive planetesimal or several
planetesimals of lower mass.

At both moderate and high resolution we observe the close approach and merging
of gravitationally bound clumps. Concerns remain about whether these
collisions are real, since our particle-mesh self-gravity algorithm prevents
bound clumps from being smaller than a grid cell. Thus the collisional cross
section is artificially large. Two observations nevertheless indicate that the
collisions can be real: we observe planetesimal mergers at both moderate and
high resolution and we see that the environment in which planetesimals merge is
rich in unbound particles. Dynamical friction may thus play an important
dissipative role in the dynamics and the merging. At high resolution we clearly
see a trailing spiral arm exerting a negative torque on a planetesimal born in
the vicinity of another planetesimal.

If the transport of newly born planetesimals into each other's Hill spheres is
physical (i.e.\ moderated by dynamical friction rather than artificial
enlargement of planetesimals and numerical viscosity), then that can lead to
both mergers and production of binaries. \cite{Nesvorny+etal2010} recently
showed that gravitationally contracting clumps of particles can form wide
separation binaries for a range of initial masses and clump rotations and that
the properties of the binary orbits are in good agreement with observed Kuiper
belt object binaries.

In future simulations strongly bound clusters of particles should be turned
into single gravitating sink particles, in order to prevent planetesimals from
having artificially large sizes. In the present paper we decided to avoid using
sink particles because we wanted to evolve the system in its purest state with
as few assumptions as possible. The disadvantage is that the particle clumps
become difficult to evolve numerically and hard to load balance. Using sink
particles will thus also allow a longer time evolution of the system and use of
proper friction times of large bodies.

The measured $\alpha$-value of MRI turbulence at $512^3$ is
$\alpha\approx0.003$. At a sound speed of $c_{\rm s}=500$ m/s, the expected
collision speed of marginally coupled m-sized boulders, based empirically on
the measurements of J07, is $\sim$$\sqrt{\alpha}c_{\rm s}\approx$ 25 -- 30 m/s.
J07 showed that the actual collision speeds can be a factor of a few lower,
because the particle layer damps MRI turbulence locally. In general boulders
are expected to shatter when they collide at 10 m/s or higher
\citep{Benz2000}.  Much larger km-sized bodies are equally prone to
fragmentation as random gravitational torques exerted by the turbulent gas
excite relative speeds higher than the gravitational escape speed
\citep{Ida+etal2008,LeinhardtStewart2009}. A good environment for building
planetesimals from boulders may require $\alpha\lesssim0.001$, as in J07.
\cite{Johansen+etal2009b} presented simulations with no MRI turbulence where
turbulence and particle clumping is driven by the streaming instability
\citep{YoudinGoodman2005}. They found typical collision speeds as low as a few
meters per second.

A second reason to prefer weak turbulence is the amount of mass available in
the disc. If we apply our results to $r=5\,{\rm AU}$, then our dimensionless
gravity parameter corresponds to a gas column density of $\varSigma_{\rm
gas}\approx 1800\,{\rm g\,cm^{-2}}$, more than ten times higher than the
Minimum Mass Solar Nebula \citep{Weidenschilling1977b,Hayashi1981}. Turbulence
driven by streaming and Kelvin-Helmholtz instabilities can form planetesimals
for column densities comparable to the Minimum Mass Solar Nebula
\citep{Johansen+etal2009b}.

The saturation of the magnetorotational instability is influenced by both the
mean magnetic field and small scale dissipation parameters, and the actual
saturation level in protoplanetary discs is still unknown. Our results show
that planetesimal formation by clumping and self-gravity benefits overall from
weaker MRI turbulence with $\alpha\lesssim0.001$. Future improvements in our
understanding of protoplanetary disc turbulence will be needed to explore
whether such a relatively low level of MRI turbulence is the case in the entire
disc or only in smaller regions where the resistivity is high or the mean
magnetic field is weak.

\begin{acknowledgements}

This project was made possible through a computing grant for five rack months
at the Jugene BlueGene/P supercomputer at J\"ulich Supercomputing Centre. Each
rack contains 4096 cores, giving a total computing grant of approximately 15
million core hours. AJ was supported by the Netherlands Organization for
Scientific Research (NWO) through Veni grant 639.041.922 and Vidi grant
639.042.607 during part of the project. We are grateful to Tristen Hayfield for
discussions on particle load balancing and to Xuening Bai and Andrew Youdin
for helpful comments. The anonymous referee is thanked for an insightful
referee report. HK has been supported in part by the Deutsche
Forschungsgemeinschaft DFG through grant DFG Forschergruppe 759 ``The Formation
of Planets: The Critical First Growth Phase''.

\end{acknowledgements}

\end{document}